\documentclass[11pt]{article}
\pdfoutput=1
\usepackage{putex}
\usepackage{amsmath,hyperref,amssymb,amsfonts,bm,amscd,slashed}
\usepackage{cite}
\usepackage{graphicx}
\usepackage{appendix}
\usepackage{array}
\usepackage{url}
\usepackage{float}
\usepackage{mathtools}
\usepackage{mathrsfs}
\usepackage{bbm}

\numberwithin{equation}{section}

\newcommand {\be} {\begin {equation}}
\newcommand {\ee} {\end {equation}}

\newcommand {\bes} {\begin {equation*}}
\newcommand {\ees} {\end {equation*}}

\newcommand{\Z}{\mathbb{Z}}

\newcommand{\R}{\mathbb{R}}
\newcommand{\C}{\mathbb{C}}

\newcommand{\beq}{\begin{equation}}
\newcommand{\eeq}{\end{equation}}

\def\<{\langle}
\def\>{\rangle}

\newcommand{\cA}{\ensuremath{\mathcal{A}}}

\newcommand{\cC}{\ensuremath{\mathcal{C}}}
\newcommand{\cD}{\ensuremath{\mathcal{D}}}

\newcommand{\cF}{\ensuremath{\mathcal{F}}}

\newcommand{\cH}{\ensuremath{\mathcal{H}}}

\newcommand{\cL}{\ensuremath{\mathcal{L}}}
\newcommand{\cM}{\ensuremath{\mathcal{M}}}
\newcommand{\cN}{\ensuremath{\mathcal{N}}}
\newcommand{\cO}{\ensuremath{\mathcal{O}}}

\newcommand{\cQ}{\ensuremath{\mathcal{Q}}}

\newcommand{\cS}{\ensuremath{\mathcal{S}}}

\newcommand{\cV}{\ensuremath{\mathcal{V}}}

\newcommand{\dd}{\mathrm{d}}

\newcommand{\zhu}{{\rm Zhu}}

\def\rQ{{\mathbbmtt{Q}\,}}

\renewcommand{\bar}{\overline}
\renewcommand{\hat}{\widehat}

\numberwithin{equation}{section}
\newcommand{\ii}{\mathrm{i}}

\DeclareMathOperator{\trace}{Tr}

\usepackage{amsthm}

\theoremstyle{plain}

\newtheorem{proposition}{Proposition}[section]

\newtheorem{remark}{Remark}[section]
\newtheorem{conjecture}{Conjecture}[section]

\theoremstyle{definition}
\newtheorem{definition}{Definition}[section]

\begin{document}
\preprint{CALT-TH 2019-040}

\institution{SCGP}{Simons Center for Geometry and Physics,	Stony Brook University,\cr Stony Brook NY 11794 USA}
\institution{Caltech}{Walter Burke Institute for Theoretical Physics, California Institute of Technology, \cr Pasadena, CA 91125, USA}

\title{From VOAs to short star products in SCFT}
\authors{Mykola Dedushenko\worksat{\SCGP,\Caltech}}

\abstract{
We build a bridge between two algebraic structures in SCFT: a VOA in the Schur sector of 4d $\cN=2$ theories and an associative algebra in the Higgs sector of 3d $\cN=4$. The natural setting is a 4d $\cN=2$ SCFT placed on $S^3\times S^1$: by sending the radius of $S^1$ to zero, we recover the 3d $\cN=4$ theory, and the corresponding VOA on the torus degenerates to the associative algebra on the circle. We prove that: 1) the Higgs branch operators remain in the cohomology; 2) all the Schur operators of the non-Higgs type are lifted by line operators wrapped on the $S^1$; 3) no new cohomology classes are added. We show that the algebra in 3d is given by the quotient $\cA_H = \zhu_{s}(V)/N$, where $\zhu_{s}(V)$ is the non-commutative Zhu algebra of the VOA $V$ (for ${s}\in{\rm Aut}(V)$), and $N$ is a certain ideal. This ideal is the null space of the (${s}$-twisted) trace map $T_{s}: \zhu_{s}(V) \to \C$ determined by the torus 1-point function in the high temperature (or small complex structure) limit. It therefore equips $\cA_H$ with a non-degenerate (twisted) trace, leading to a short star-product according to the recent results of Etingof and Stryker. The map $T_{s}$ is easy to determine for unitary VOAs, but has a much subtler structure for non-unitary and non-$C_2$-cofinite VOAs of our interest. We comment on relation to the Beem-Rastelli conjecture on the Higgs branch and the associated variety. A companion paper will explore further details, examples, and some applications of these ideas.
}

\date{}

\maketitle

\tableofcontents
\setlength{\unitlength}{1mm}

\newpage

\section{Introduction}

The discovery of correspondence between 4d $\cN=2$ superconformal field theories (SCFTs) and 2d vertex operator superalgebras (VOSAs or VOAs for short) in \cite{Beem:2013sza} has fueled a lot of work in the past six years, including constructions in other dimensions \cite{Beem:2014kka,Chester:2014mea,Beem:2016cbd,Dedushenko:2016jxl,Dedushenko:2017avn,Dedushenko:2018icp,Mezei:2018url}. This ``SCFT/VOA'' or ``4d/2d'' relation differs from other types of 4d/2d correspondences (such as those in \cite{Nekrasov:2015wsu,Nekrasov:2017gzb,Alday:2009aq,Gadde:2013sca,Dedushenko:2017tdw,Feigin:2018bkf}) in that the VOA structure is very explicitly identified with certain OPE coefficients in the parent 4d SCFT. The constructions in other dimensions motivated by the SCFT/VOA correspondence include: protected VOAs in 6d $(2,0)$ theories \cite{Beem:2014kka}, associative algebra (1d TQFT) in 3d $\cN=4$ theories \cite{Chester:2014mea,Beem:2016cbd}, and 3d TQFT sector of the 5d maximal super Yang-Mills \cite{Mezei:2018url}. Star-products appearing in 3d are also the subject of mathematical study in \cite{ERS1, ERS2}, and a recent bootstrap-related work \cite{Chang:2019dzt} .

The two most studied constructions are in four and three dimensions\footnote{See recent works \cite{Agmon:2017xes,Chester:2018aca,Agmon:2019imm} for nice applications of the 3d construction.}, and the goal of this paper is to connect them via the Kaluza-Klein (KK) reduction on the circle. However, the original formulation of \cite{Beem:2013sza,Chester:2014mea,Beem:2016cbd} does not admit this, as it uses conformal invariance both in 4d and in 3d, which dimensional reduction explicitly breaks. Therefore we need an alternative formulation that does not rely on conformal symmetry. Physically, the works \cite{Dedushenko:2016jxl,Dedushenko:2017avn,Dedushenko:2018icp} have formulated the construction of 1d sector for general 3d $\cN=4$ theories away from conformal points by placing them on $S^3$, and developed an extensive set of tools to compute associative algebras (and 1d correlators) associated to a large class of Lagrangian 3d $\cN=4$ theories (which is yet to be extended to include the most general Lagrangian 3d $\cN=4$ QFTs).

In 4d $\cN=2$ case, the same has been done recently in \cite{Pan:2019bor, Dedushenko:2019yiw} by placing the theory on $S^3\times S^1$. Unlike in 3d, the 4d case still requires the theory to be conformal: even though the construction on $S^3\times S^1$ does not refer to superconformal symmetry \emph{explicitly}, it uses the $U(1)_r$ R-symmetry, which is only present in SCFTs (this can be proved using the submultiplet of anomalies of the multiplet of currents). Therefore the scope of Lagrangian tools of \cite{Pan:2019bor,Dedushenko:2019yiw} is more limited than in 3d, and only applies to theories whose matter content satisfies the conformality constraint. At the same time in \cite{Oh:2019bgz,Jeong:2019pzg}, an alternative formulation based on a version of the Omega-background \cite{Nekrasov:2002qd,Nekrasov:2003rj,Nekrasov:2010ka} was described, which has the same property: though it does not refer to conformal symmetries explicitly, it uses the $U(1)_r$, thus only applies to SCFTs.  

The $S^3\times S^1$ formulation can be readily applied to general non-Lagrangian 4d $\cN=2$ SCFTs, while the Omega-background formulation, though probably exists, does not seem accessible for non-Lagrangian theories. This, together with some other aspects that will become transparent later, makes the $S^3\times S^1$ background the most handy for us: it preserves the VOA construction and, by shrinking the size of $S^1$, interpolates between the 2d and 1d protected sectors. Besides the references mentioned so far, there has been a huge number of works on the SCFT/VOA correspondence \cite{Beem:2014rza,Lemos:2014lua,Cordova:2015nma,Bobev:2015kza,Liendo:2015ofa,Buican:2015tda,Cecotti:2015lab,Lemos:2015orc,Nishinaka:2016hbw,Buican:2016arp,Xie:2016evu,Cordova:2016uwk,Arakawa:2016hkg,Bonetti:2016nma,Song:2016yfd,Creutzig:2017qyf,Fredrickson:2017yka,Cordova:2017mhb,Song:2017oew,Buican:2017fiq,Beem:2017ooy,Neitzke:2017cxz,Pan:2017zie,Fluder:2017oxm,Buican:2017rya,Choi:2017nur,Arakawa:2017fdq,Wang:2018gvb,Niarchos:2018mvl,Creutzig:2018lbc,Dedushenko:2018bpp,Bonetti:2018fqz,Arakawa:2018egx,Nishinaka:2018zwq,Agarwal:2018zqi,Beem:2018duj,Costello:2018zrm,Buican:2019huq,Xie:2019yds,Beem:2019tfp,Xie:2019zlb,Fluder:2019dpf,Watanabe:2019ssf}, both in physics and math, and some of their results will prove helpful along the way.

In the current paper, we connect the constructions in 4d and 3d in cases when the 3d $\cN=4$ theory appears as the Kaluza-Klein (KK) reduction of the 4d $\cN=2$ theory. The KK reduction of 4d $\cN=2$ has received attention in the past, \emph{e.g.}, \cite{Seiberg:1996nz} primarily focused on the Coulomb branch sector, \cite{Dolan:2011rp,Gadde:2011ia,Imamura:2011uw,Aharony:2013dha} studied indices and dualities, and \cite{Buican:2015hsa,Agarwal:2018oxb} explored KK reduction of Argyres-Douglas theories. Here we are interested in a somewhat intricate question of dimensionally reducing the whole VOA protected sector.

Through very similar in structure, the 4d and 3d SCFT constructions of \cite{Beem:2013sza,Chester:2014mea,Beem:2016cbd} (we call them the ``$Q+S$'' constructions) explicitly use conformal invariance. Reducing a theory on the circle completely breaks conformal symmetry and induces a renormalization group (RG) flow that lands at some 3d $\cN=4$ SCFT in the end. Only at the UV and IR endpoints of the RG flow, the ``$Q+S$'' constructions are available, but not along the flow, which naively suggests that we have no controllable way to relate them.

The resolution, as should be clear from the above discussion, is to consider 4d $\cN=2$ theories on $S^3\times S^1$ instead, where $S^3$ has radius $\ell$ and $S^1$ has circumference $\beta\ell$. Much like the Omega-background, we can think of it as a deformation (parametrized by  $\ell^{-1}$) of the flat space theory. Furthermore, because the cylinder $S^3\times \R$ is conformally flat, there exists a canonical way to put an SCFT on it, which is standard in the radial quantization. Imposing certain (twisted) periodic supersymmetric boundary conditions along $\R$, one then easily replaces it by $S^1$. For locality reasons, it is natural to expect that in the $\beta\to0$ limit, the theory on $S^3\times S^1$ reduces to the $S^3$ version (or the $S^3$ deformation) of the same 3d theory one obtains via the KK reduction in flat space, \emph{i.e.} on $\R^3\times S^1$.

As was explained in \cite{Dedushenko:2019yiw}, and as we briefly review below, the chiral algebra construction survives on $S^3\times S^1$, and it only uses the ordinary isometries of this background. This being so, it exists at arbitrary value of $\beta$, thus providing the missing link between the UV and IR fixed points. At any finite value of $\beta$, the construction of \cite{Dedushenko:2019yiw} gives a VOA (of course the same as in \cite{Beem:2013sza}) on the two-torus embedded in $S^3\times S^1$, and in the $\beta\to0$ limit, the 3d construction should emerge. In this limit, the torus degenerates to a circle, so we should study the ``small complex structure limit'' ($\tau\to0$, where $\tau=i\beta/(2\pi)$) of torus correlators. We also refer to this as the ``high temperature limit'' for obvious reasons. In this limit, as we will see, the VOA structure indeed encodes a certain associative algebra equipped with the trace map, which is what describes the 1d protected sector of the 3d theory.

In fact, there exists a natural notion of ``dimensional reduction'' of a VOA $V$ in the mathematics literature, -- the associative algebra $\zhu(V)$ \cite{Zhu} known as the Zhu algebra.\footnote{Even though implicit in numerous works, to the best of authors knowledge, it has never been acknowledged explicitly that the Zhu algebra can be thought of as the dimensional reduction of a vertex algebra.} This algebra is highly important as it ``controls'' the representation theory of the VOA: it turns out that the VOA structure is so rigid that it is enough to find a module over the associative algebra $\zhu(V)$, and then one can uniquely (up to isomorphism) extend it to the full $V$-module. One could naively guess that this algebra is what we need -- that $\zhu(V)$ is precisely what arises in the $\beta\to0$ limit. This turns out to be close to the answer but not completely true. One reason is that we should consider a slightly more general notion of an algebra $\zhu_g(V)$, associated to a finite order automorphism $g$ of a vertex operator superalgebra $V$, that controls the $g$-twisted representation theory of $V$ \cite{Dong1998,Dong_Zhao_06}. (We will need a particular automorphism $g={s}$, in which case $\zhu_{s}(V)$ also appears as $\zhu_H V$ in \cite{DSK}.) Another reason is that this $\zhu_{s}(V)$ is too large and contains many operators that are ``lifted'' in the 3d limit via an interesting mechanism involving line operators. In short, this is because $V$ contains Schur operators with spin (or ``spinning'' Schur operators), which do not survive the 3d limit (in 3d, only scalar operators belong to the protected sector). Many of these operators contribute non-trivially to $\zhu_{s}(V)$, thus making it too large. Such operators form an ideal $N\subset \zhu_{s}(V)$, and the correct answer for the 3d algebra is $\zhu_{s}(V)/N$.

As we find, another way to see it is from the $\tau\to 0$ limit of torus correlators, in particular the torus one-point function. We can use it to construct an ${s}$-twisted trace on $\zhu_{s}(V)$,
\begin{equation}
T_{s}: \zhu_{s}(V) \to \C,
\end{equation}
which can further be applied to compute correlators of operators in $\zhu_{s}(V)$,
\begin{equation}
\langle \alpha_1\star \dots \star \alpha_n\rangle \equiv T_{s}(\alpha_1\star \dots \star \alpha_n).
\end{equation}
This trace, however, is degenerate, meaning that the form $B_T(\alpha,\beta) = T_{s}(\alpha\star\beta)$ (here $\star$ is the product in $\zhu_{s}(V)$) has a kernel, which is precisely what we called $N$ before. This kernel is obviously a two-sided ideal, and we can consider a quotient $\cA_H=\zhu_{s}(V)/N$, now equipped with a non-degenerate ${s}$-twisted trace,
\begin{equation}
T_{s}: \cA_H=\zhu_{s}(V)/N \to \C.
\end{equation}
The unitarity in 3d demands that the 1d protected associative algebra be equipped with a non-degenerate ${s}$-twisted trace \cite{Beem:2016cbd}, and we argue that $(\cA_H, T_{s})$ is precisely this data. The work of \cite{ERS1, ERS2} proves that such traces are in one-to-one correspondence with the short star-products. Shortness is another name for the ``truncation condition'' introduced in \cite{Beem:2016cbd}, and is a distinguishing property of star-products that appear in 3d $\cN=4$ theories via the construction of \cite{Chester:2014mea,Beem:2016cbd}.

One possible caveat of the above discussion is that the star-product we find is not necessarily the one corresponding to the SCFT point in 3d. Instead, it might be its deformation. Indeed, \cite{ERS1} show that there exists a finite-dimensional family of star products, so in principle, we could have landed at any of them in the 3d limit. In a companion paper \cite{DW}, we distinguish cases where we actually get the SCFT star-products from those where we obtain deformations thereof. In short, the criterion is as follows: whenever the $U(1)_r$ symmetry of the 4d theory directly enhances to the $SU(2)_C$ R-symmetry in 3d, we get the SCFT star-product; if the $U(1)_r \rightsquigarrow SU(2)_C$ enhancement involves mixing with the (Coulomb branch) flavor symmetries, we obtain a deformation of the SCFT star product by the imaginary FI terms (see \cite{Buican:2015hsa}). 

Finally, this is closely related to the question of determining the Higgs branch of either the 4d or 3d theory. The algebra $\cA_H$ is a filtered quantization of the Higgs branch, thus it encodes its complex geometry in either three or four dimensions. The commutative limit, ${\rm gr }\cA_H = {\rm gr}\zhu_{s}(V) / {\rm gr} N$, is the Higgs branch chiral ring, commonly assumed to be the ring of regular functions on the Higgs branch, and so it is natural to inquire about the relation to associated variety \cite{Arakawa2012}, which is the content of the Beem-Rastelli conjecture \cite{Beem:2017ooy}. We make some comments about this towards the end.

This paper has the following structure. We start in Section \ref{prelim} with some generalities about SCFTs on $S^3\times \R$, and discuss the cylinder counterpart of the VOA construction of \cite{Beem:2013sza}. In Section \ref{reduction_phys} we discuss the physics of dimensional reduction: explain necessary details on how the 4d theory on $S^3\times S^1$ reduces to the 3d theory in the small-circle limit, and how line operators remove some local operators from the cohomology in the 3d limit. Section \ref{sec:Zhu} gives motivation for the appearance of Zhu algebra in dimensional reduction, and provides a mathematical review of the (twisted) Zhu algebra construction. Section \ref{sec:trace} discusses torus correlators of the VOA, their high-temperature limit, modularity, and how they encode twisted traces on the Zhu algebra, which is at the heart of our construction. In Section \ref{sec:comments}, we briefly illustrate this for affine VOAs, and also mention relation to the $C_2$ algebra. Finally, we conclude in Section \ref{conclude} with the outline of future directions.

\emph{Warning:} in this paper, the algebras are generally $\frac12\Z$-graded, which is slightly more natural physically, but differs from the conventions in \cite{ERS1,Beem:2016cbd}, where the $\Z$-grading was adopted.

\section{SCFT on $S^3\times \R$ and $S^3\times S^1$}\label{prelim}
\subsection{Generalities}
A 4d CFT can be placed on a cylinder $S^3\times\R$ via the Weyl transformation from the flat space. For general 4d CFT, this introduces an ambiguity in the vacuum (Casimir) energy due to finite local counterterms, however this procedure is unambiguous for SCFTs \cite{Assel:2014tba,Assel:2015nca,Bobev:2015kza}. Anomalous Weyl transformation of the stress-energy tensor generates the Casimir energy on $S^3$, which for round $S^3$ of radius $\ell$ is known to be \cite{Assel:2015nca}
\begin{equation}
E_0 = \frac{4}{27\ell}(a+3c),
\end{equation}
where $a$ and $c$ are Weyl anomaly coefficients in a 4d SCFT. We will further consider closing the $\R$ direction into a circle $S^1$ with supersymmetric boundary conditions.

Let us for now focus on 4d $\cN=2$ SCFTs in the Euclidean space $\R^4$. The flat space coordinates $(x^1, x^2, x^3, x^4)$ are organized into complex coordinates,
\begin{equation}
z = x^3 + \ii x^4,\quad z'=x^1 + \ii x^2\,.
\end{equation}
We also introduce spherical coordinates $(r,\theta,\varphi,\tau)$ in $\R^4$, where $(\theta,\varphi,\tau)$ are fibration coordinates on $S^3$. In terms of these,
\begin{equation}
z=r\sin\theta e^{\ii\varphi},\quad z'=r\cos\theta e^{\ii\tau}.
\end{equation}
A change of coordinates
\begin{equation}
\label{ry_coord}
r= \ell e^{y/\ell},
\end{equation}
defines $(\theta,\varphi,\tau,y)$, -- coordinates on $S^3\times\R$ adapted to the Weyl transformation
\begin{equation}
\label{weyl}
\dd s^2_{\R^4} = \dd r^2 + r^2 \dd\Omega^2 = e^{2y/\ell}\left( \dd y^2 + \ell^2\dd\Omega^2 \right) \longrightarrow \dd s^2_{S^3\times \R}= \dd y^2 + \ell^2\dd\Omega^2,
\end{equation}
relating the flat metric to the cylinder metric. As in \cite{Dedushenko:2019yiw}, we can observe how this procedure affects the superconformal symmetry of the flat space theory. While \cite{Dedushenko:2019yiw} was specifically concerned with theories admitting Lagrangian description, it is true in full generality that superconformal transformations are parametrized in terms of conformal Killing spinors $\xi_{A\alpha}$, $\bar{\xi}_A^{\dot{\alpha}}$, which in the flat space take the simplest form,
\begin{equation}
\xi_A = \epsilon_A + i\slashed{x}\eta_A\,,\quad \bar{\xi}_A = \bar{\epsilon}_A + i\slashed{x}\bar{\eta}_A\, .
\end{equation}
Upon the Weyl transformation \eqref{weyl}, the frame is rescaled by $e^{-y/\ell}$, and the spinors are rescaled by $e^{-y/(2\ell)}$. On the other hand, the explicit factor of $\slashed{x}$ scales as $e^{y/\ell}$ far away from the origin. We conclude that half of the conformal Killing spinors (those parametrized by $\epsilon_A$ and $\bar{\epsilon}_A$) behave as $e^{-y/(2\ell)}$, and half (those parametrized by $\eta_A$ and $\bar{\eta}_A$) are proportional to $e^{y/(2\ell)}$ on $S^3\times\R$.

Imposing the simple periodic identification along $\R$, \emph{i.e.} replacing it by $S^1$, 
\begin{equation}
\label{periodic}
y \sim y+ \beta\ell,
\end{equation}
breaks supersymmetry completely: the above observation on $y$-dependence shows that the conformal Killing spinors are simply not periodic in $y$. To preserve at most half of the supercharges on $S^3\times S^1$, we should consider our theory in the twisted sector with respect to the $SU(2)_R$ R-symmetry \cite{Dedushenko:2019yiw}. What this means is that for any variable $\cF$ in the theory (either a field or a SUSY parameter), we impose\footnote{This is of course equivalent to turning on an imaginary background connection for the R-symmetry.}
\begin{equation}
\label{tw_sector}
\cF(y+\beta\ell) = e^{-\beta R}\cF(y),
\end{equation}
where $R$ is the $SU(2)_R$ charge of $\cF$. This indeed preserves half of the supercharges, namely those for which the failure of periodicity in $y$ coincides with the one dictated by their R-charge. Clearly, these are the supercharges that commute with $\cH-R$, where the dilatation operator $\cH$ generates translations in the $y$ direction on the cylinder. Therefore the algebra of symmetries on $S^3\times S^1$ is given by the centralizer of $\cH-R$ inside of the full 4d $\cN=2$ superconformal algebra $\mathfrak{su}(4|2)$.

Using the conventions of \cite{Dedushenko:2019yiw}, one identifies the algebra of symmetries $\mathfrak{s}$ on $S^3\times S^1$ as the central extension of $\mathfrak{su}(2|1)_\ell\oplus \mathfrak{su}(2|1)_r$, with the central charge being the $\cH-R$ itself. Using the 4d $\cN=2$ (anti-)commutation relations in the conventions of \cite{Beem:2013sza}, we identify the subalgebra $\mathfrak{s}$ explicitly, with the most important anti-commutation relation given by
\begin{align}
\{\cQ_\alpha^1, \cS_1^\beta\} & \ = \ \frac12 \delta_\alpha^\beta (\cH-R) + \cM_\alpha{}^\beta -\delta_\alpha^\beta\,\frac{r+R}2\,,\\
\{\tilde{\cS}^{2\dot{\alpha}},\tilde{\cQ}_{2\dot{\beta}}\} & \ = \ \frac12 \delta^{\dot{\alpha}}_{\dot{\beta}}(\cH-R) +\cM^{\dot \alpha}{}_{\dot \beta} + \delta^{\dot \alpha}_{\dot{\beta}}\,\frac{r-R}2\,.
\end{align}
Here the first relation is the anti-commutator in the left $\mathfrak{su}(2|1)$ that contains left rotations $\cM_\alpha{}^\beta$, while the second line corresponds to the right $\mathfrak{su}(2|1)$ and right rotations $\cM^{\dot\alpha}{}_{\dot\beta}$.

\subsection{Chiral algebra and Schur operators on $S^3\times S^1$}
The two supercharges relevant for the chiral algebra construction of \cite{Beem:2013sza},
\begin{align}
\rQ_1& \ = \  \cQ^1_- + \frac1{\ell}\tilde{\cS}^{2\dot -}\,,\cr
\rQ_2& \ = \ \frac1{\ell}\cS_1^- - \tilde{\cQ}_{2\dot -}\,,
\end{align}
belong to the algebra of symmetries on $S^3\times S^1$. Therefore the chiral algebra construction naturally exists on $S^3\times S^1$, as was explained in details in \cite{Dedushenko:2019yiw}, and as we will see, this background is perfectly tailored for our needs. The operators in the algebra are, as usual, defined through the cohomology of $\rQ_{1,2}$. In the flat space case of \cite{Beem:2013sza}, the cohomology classes formed a vertex operator algebra (VOA) living on the plane $\C\subset \C^2$ parametrized by $z$. Upon Weyl transformation, the chiral algebra plane becomes a cylinder $S^1_\varphi\times \R \subset S^3\times\R$, which turns into a torus after the periodic identification of $\R$ discussed above. The torus is embedded into the spacetime as $S^1_\varphi\times S^1 \subset S^3\times S^1$, where $S^1_\varphi\subset S^3$ is a great circle at $\theta=\pi/2$, the location at which the $\tau$ circle collapses to a point. We introduce a complex coordinate on the cylinder/torus,
\begin{equation}
w=\ell\varphi-iy,
\end{equation}
which is related to the flat space coordinate on the chiral algebra plane by
\begin{equation}
z=\ell e^{iw/\ell}.
\end{equation}

The cohomology classes are represented by the twisted-translated Schur operators in flat space \cite{Beem:2013sza}, and we can easily identify their counterpart on $S^3 \times \R$ through the Weyl transformation that relates it to $\R^4$. We start by giving the answer, and then explain it. So, the twisted-translated operators on $\R^4$ and $S^3\times\R$ respectively are defined by
\begin{align}
\label{tw_tr}
\cO^{\rm flat}(z)&=\cO^{\rm flat}_{a_1\dots a_n}(z,\bar{z}, z'=\bar{z}'=0) u^{a_1}(\bar{z})\dots u^{a_n}(\bar{z}),\quad \text{where } u^a=(1,\bar{z}/\ell),\cr
\cO^{\rm cyl}(w) &= \cO^{\rm cyl}_{a_1\dots a_n}(w,\bar{w},\theta=\pi/2)u^{a_1}(\bar{w})\dots u^{a_n}(\bar{w}),\quad \text{where } u^a(\bar w)=i^{1/2}(e^{i\bar{w}/(2\ell)}, e^{-i\bar{w}/(2\ell)}).\cr
\end{align}
While the former is the usual definition from \cite{Beem:2013sza}, the latter is what needs an explanation.

 To that end, we first recall how a scalar (quasi-)primary observable $\Phi$ of conformal dimension $E$ scales under the Weyl transformations $\dd s^2_{\R^4} \mapsto \dd s^2_{S^3\times\R}=e^{-2y/\ell}\dd s^2_{\R^4}$. Namely, the flat space observable $\Phi_{\rm flat}$ and the cylinder observable $\Phi_{\rm cyl}$ are related by
\begin{equation}
\label{weylOp}
\Phi_{\rm flat}= \left(e^{y/\ell} \right)^{-E} \Phi_{\rm cyl}.
\end{equation}
Spinning observables that are still primary scale in the same way, but because their definition in general requires the choice of local frame, we might accompany Weyl transformation by a local frame rotation. This is what happens in our case, and while in the flat space we work with the standard translation-invariant frame,
\begin{align}
e^i &= \dd x^i\,,\quad i=1\dots 4,
\end{align}
on the cylinder $S^3\times \R$, we prefer to work with the frame that behaves well under isometries of this space. Two isometries which are important to us are $\partial_y$ (translations along $\R$) and $\partial_\varphi$ (a $U(1)$ isometry of $S^3$), which generate translations of the chiral algebra torus $S^1_\varphi\times S^1 \subset S^3\times S^1$. The flat space frame is not invariant under these translations, but if we rotate $e^{3,4}$ by an angle $\varphi+\frac{\pi}{2}$ (while keeping $e^{1,2}$ intact), we obtain a frame invariant under these two isometries. This extra frame rotation amounts to the following modification of the transformation law \eqref{weylOp}, which now includes both Lorentz and Weyl transformations,
\begin{equation}
\Phi_{\rm flat}= e^{-i(j_1+j_2)(\varphi+\pi/2)} \left(e^{y/\ell} \right)^{-E} \Phi_{\rm cyl},
\end{equation}
where $j_1$ and $j_2$ are the left and right spins of $\Phi$ respectively.

Now we would like to understand how twisted-translated Schur operators behave under the Weyl transformation. The scalar Schur operators $\cO^{\rm flat}_{1\dots 1}$ are given by the $SU(2)_R$ highest-weight components of the Higgs branch operators $\cO^{\rm flat}_{a_1\dots a_n}$ of R-charge $R=n/2$ and conformal dimension $E=2R=n$, and they are known to be primary in 4d \cite{Beem:2013sza}. Thus the Weyl transformation is simple,
\begin{equation}
\label{weyl_higgs}
\cO^{\rm flat}_{a_1\dots a_n} = (e^{y/\ell})^{-E} \cO^{\rm cyl}_{a_1\dots a_n},
\end{equation} 
and implies the following relation between the observables defined in \eqref{tw_tr},
\begin{equation}
\label{weyl_2d}
\cO^{\rm flat}(z)= \left(iz/\ell\right)^{-h} \cO^{\rm cyl}(w),
\end{equation}
thereby justifying the definition of $\cO^{\rm cyl}(w)$ in \eqref{tw_tr}. Indeed, this shows that the Weyl image of $\cO^{\rm flat}(z)$ is $\cO^{\rm cyl}(w)$. Furthermore, equation \eqref{weyl_2d} is precisely how primary vertex operators of dimension $h$ transform under the conformal map $z=\ell e^{iw/\ell}$ in 2d.

The rest of Schur operators, however, are neither scalar nor superconformal primary in 4d: the Schur operators of types $\cD_{R(0,j_2)}$, $\bar\cD_{R(j_1,0)}$, and $\hat\cC_{R(j_1,j_2)}$ are given by
\begin{equation}
\widetilde{\cQ}_{\dot+}^1 \Psi\,,\quad {\cQ}_{+}^1 \Psi\,,\quad \cQ_+^1\widetilde{\cQ}_{\dot+}^1 \Psi\,
\end{equation}
respectively, where $\Psi$ denotes the appropriate superconformal primaries \cite{Beem:2013sza}. In general, their transformation rules would be determined through derivatives and super-derivatives of those of primary operators. Taking into account the frame rotation as mentioned previously, the Weyl transformation of a general Schur operator is
\begin{equation}
\cO^{\rm flat}_{a_1\dots a_n} = i^{-j_1-j_2}e^{-i(j_1+j_2)\varphi}(e^{y/\ell})^{-E} \cO^{\rm cyl}_{a_1\dots a_n} + \dots,
\end{equation}
whereth the ellipses represent extra terms resulting from the ``non-primarity'' of $\cO$. Notice that these more general Schur operators have
\begin{equation}
E=2R+j_1+j_2,
\end{equation}
and we find that the twisted-translated operators transform as
\begin{equation}
\label{weyl_2d_gen}
\cO^{\rm flat}(z)= \left(iz/\ell\right)^{-h} \cO^{\rm cyl}(w) + \dots,
\end{equation}
where the ellipsis again represent corrections due to $\cO$ being a descendant, and the 2d conformal dimension is now \cite{Beem:2013sza}
\begin{equation}
h=R+j_1+j_2.
\end{equation}
In fact, the missing terms written as $\dots$ above are uniquely fixed by conformal invariance (recall that the VOAs that appear in this context are conformal).

Therefore, at the end of the day we arrive to a somewhat obvious (and definitely expected) conclusion: the VOA living on $S^1_\varphi\times\R\subset S^3\times\R$ can be obtained from the flat space VOA by the usual conformal map $z/\ell=e^{iw/\ell}$. Passing to the torus $S^1_\varphi\times S^1\subset S^3\times S^1$ then simply corresponds to imposing periodic identification \eqref{periodic} along $\R$ in the VOA.

Let us spell out the boundary (or rather periodicity) conditions in the VOA on $S^1_\varphi\times S^1$. Periodicity along $S^1_\varphi$ is induced in the usual way by the conformal map $z/\ell=e^{iw/\ell}$ -- since vertex operators were single-valued on $\C_z$, they are defined in the NS sector on $S^1_\varphi \times \R$. In other words, integer spin operators are periodic, and half-integer spin operators are anti-periodic along $S^1_\varphi$ (recall that there is no relation between spin and statistics in these VOAs). We refer to this by saying that $S^1_\varphi$ has the NS spin structure. As for the other circle, the twisted sector periodicity \eqref{tw_sector} of the 4d fields and the definition of $\cO^{\rm cyl}(w)$ in \eqref{tw_tr} imply that all the 2d vertex operators are  periodic along it. We refer to this by saying that the $S^1$ has the R spin structure, and the torus $S^1_\varphi\times S^1$ has the NS-R spin structure.

Thus our basic setup is the half-integer graded VOSA on the torus with NS-R spin-structure. Recall from \cite{Dedushenko:2019yiw} that it is possible to obtain other spin structures as well, by turning on a $(-1)^{2(R+r)}$ monodromy around the $S^1$ in the geometry $S^3\times S^1$ and by inserting a surface defect along $S^1_\tau\times S^1$, where $S^1_\tau\subset S^3$ is a circle at $\theta=0$ parametrized by $\tau$. Upon the KK reduction along $S^1$, the surface defect becomes a certain R-symmetry vortex line on $S^3$, whereas the monodromy corresponds to the twisted KK reduction along $S^1$ (for example, in Lagrangian theories, this $(-1)^{2(R+r)}$ does not affect vector multiplets, while hypers acquire an additional minus sign upon going around the $S^1$). These generalizations provide interesting future directions, especially the monodromy one (which results in the NS-NS spin structure on the torus), as in this case the KK reduction will produce a different 3d theory. However, even the basic case of NS-R spin structure, with no monodromies or defects added, is interesting enough, hence we limit ourselves to such a situation. 

\section{Physics of the 4d$\to$ 3d reduction}\label{reduction_phys}

We would like to study the $\beta\to0$ limit of $\cN=2$ SCFTs on $S^3\times S^1$, \emph{i.e.}, the limit of zero circumference of $S^1$. We refer to this as the KK limit for obvious reasons. Let us contrast this to the KK reduction of the flat space $\cN=2$ SCFT. Placing a theory on $\R^3\times S^1$ completely breaks conformal (super)charges and induces an RG flow. The end point of this RG flow is some 3d $\cN=4$ SCFT, and we can think of the parent 4d SCFT as the UV completion of this 3d theory, the radius of the KK circle being the RG scale. In this picture the UV SCFT has one protected sector described by the VOA, and the IR SCFT has another protected sector described by an associative algebra. At the intermediate energy scales, these structures are lost, so there is no obvious relation between them.

Now on $S^3\times S^1$, taking the analogous IR limit corresponds to $\beta\to 0$.\footnote{Since $\beta$ is the circumference of $S^1$ divided by the radius of $S^3$, we can interpret this as putting a 4d theory on $S^3\times S^1$ with $S^3$ of extremely large radius and $S^1$ of finite radius, and then zooming out to the scales where $S^3$ is of finite radius and $S^1$ is really small.} The algebra of symmetries at finite $\beta$, namely the centrally extanded $\mathfrak{s}=\mathfrak{su}(2|1)_\ell\oplus\mathfrak{su}(2|1)_r$, is known to be the 3d $\cN=4$ SUSY algebra on $S^3$ with the central charge \cite{Dedushenko:2016jxl}. In the $\beta\to 0$ limit it enhances to the 3d superconformal algebra $\mathfrak{osp}(4|4)$. However, the main advantage of the $S^3\times S^1$ background is that the chiral algebra construction is based on the non-conformal algebra $\mathfrak{s}$, and thus exists at all radii of $S^1$. Therefore, we can interpolate between the VOA of the 4d theory and the associative algebra of the 3d theory in a controlled way.

\subsection{A Cardy-like counterterm}
Thinking of a $(d+1)$-dimensional SCFT on $S^d\times S^1$ as the UV completion of the $d$-dimensional theory on $S^d$ is quite subtle. Unlike with other UV regulators, here at the finite ``cut-off'' $\beta$, we have much more degrees of freedom, -- the whole tower of KK modes that must be integrated out in the $d$-dimensional limit. This makes the naive $\beta\to 0$ limit divergent. In fact, the $S^d\times S^1$ partition function counts states (or more specifically, BPS states) in the Hilbert space $\cH[S^d]$, with the counting parameter $q=e^{-\beta}$. If $\beta\to 0$, then $q\to 1$, and all the states are counted with the same weight, which produces divergence simply because $\cH[S^d]$ (or the space of BPS states) is infinite-dimensional. The asymptotics at $\beta\to0$ is called the Cardy behavior due to the related work \cite{CARDY1986186} in 2d CFT. This quantity can be computed \cite{DiPietro:2014bca,Buican:2015ina,Ardehali:2015bla,Chang:2019uag}, and the universal expression for a large class of 4d $\cN=2$ SCFTs \cite{Beem:2017ooy} is given by the Di Pietro-Komargodski formula,\footnote{This holds when $c_{\rm 4d}> a_{\rm 4d}$ \cite{Ardehali:2015bla}, which we assume. The case of $c_{\rm 4d}\leq a_{\rm 4d}$ will be considered elsewhere.}
\begin{equation}
\label{cardy}
\log Z[S^3\times S^1] \sim \frac{8\pi^2 (c_{\rm 4d}-a_{\rm 4d})}{\beta},
\end{equation}
where $c_{\rm 4d}$ and $a_{\rm 4d}$ are the Weyl anomalies of the 4d SCFT.

To have a finite $\beta\to0$ limit giving the $S^3$ partition function, we ought to subtract this Cardy-like divergence by a counterterm.\footnote{This might still give an infinite answer if the resulting 3d $\cN=4$ theory is ``bad'' in the terminology of \cite{Gaiotto:2008ak}, which is closely related to the 4d theory having $c_{\rm 4d}\leq a_{\rm 4d}$.} From the 3d point of view, this is interpreted as a supersymmetric counterterm linear in the UV cut-off $\Lambda= \frac1{\beta\ell}$. There exists only one such counterterm, which is the supergravity extension of the Einstein-Hilbert action \cite{Closset:2012vg,Closset:2012vp,Closset:2012ru},
\begin{equation}
\label{counterterm}
S_{\rm ct} = \Lambda\times \frac23 (c_{\rm 4d}-a_{\rm 4d})\int\dd^3x \sqrt{g} (R+\dots),
\end{equation}
where the coefficient in front of the integral was tuned to cancel \eqref{cardy}. Here the ellipsis represents SUSY completion, and its form is not relevant as it vanishes for the $S^3$ background.

Indeed, the local counterterms in 3d $\cN=4$ SUGRA can be constructed from those in 3d $\cN=2$ listed in \cite{Closset:2012vg,Closset:2012vp,Closset:2012ru}. Alternatively, one can obtain them by dimensional reduction of 4d couterterms, classified in \cite{Assel:2014tba}, on a circle of circumference $1/\Lambda$. Linearly divergent counterterms in 3d correspond to quadratic divergences in 4d, and the Einstein-Hilbert counterterm is the one with such a property, and the only one that works for us. (The FI counterterm has a wrong dependence on the radius $\ell$.)

\begin{remark}
	In fact, the linear counterterms are the only ones that are both supersymmetric and divergent. Therefore, if \eqref{cardy} included any other divergences besides the linear one, we simply would not have a supersymmetric counterterm to cancel them, thus the physically sensible 3d limit would be ill-defined. This again is related to the ``$c_{\rm 4d}\leq a_{\rm 4d}$'' theories in 4d, and ``bad'' theories in 3d.
\end{remark}

After subtracting the linear divergence, one could worry whether the finite part is unambiguous, \emph{i.e.}, whether there exist finite counterterms that could render the answer unphysical. Fortunately, this does not happen: such finite counterterms in 3d would correspond to linearly divergent counterterms in 4d, and again referring to \cite{Assel:2014tba}, there are none. So to summarize, our definition of the $S^3$ partition function is given by the $\beta\to 0$ limit of the $S^3\times S^1$ partition function (the Schur index) with the counterterm \eqref{counterterm} included.

\subsection{Observables and the cohomology}\label{lines}
Since we study algebraic structures on spaces of local operators, it is important to understand what is the relation between such spaces in 4d and 3d. Under the KK reduction, local observables in 4d obviously map to local observables in 3d. In addition, line operators in 4d can also produce local operators in 3d by wrapping\footnote{See \cite{Saberi:2019ghy} for a recent reappearance of this idea in other contexts and dimensions.} the circle $S^1$, and these operators might be genuinely new, meaning that they might not descend from local operators in 4d.

While local operators form a vector space, line operators are objects in the category. Local operators decorating lines correspond to morphisms in this category, and ordinary local operators in 4d appear as endomorphisms of the trivial line. By wrapping a line on $S^1$ and treating it as a local operator in lower dimension, we get, loosely speaking, a correspondence:
\begin{equation}
\cL=\left\{\text{Lines in 4d}\right\} \longrightarrow \cV_{\rm 3d}^{\rm loc}=\left\{\text{Local operators in 3d}\right\}.
\end{equation}
This should be thought of as part of a bigger structure that describes at once the relation between all extended objects in 4d and all extended objects in 3d. We do not attempt to study it here, but we need one more piece of it, which is the mapping between the spaces of local operators:
\begin{equation}
\label{ops_extension}
\cV_{\rm 4d}^{\rm loc}\to \cV_{\rm 3d}^{\rm loc}.
\end{equation}
This map is neither injective (operators that are derivatives along the KK circle of some other operator often decouple in the 3d limit), nor surjective, as the space of local operators can enlarge upon the reduction. The best-know example of the latter phenomenon is given by Wilson and 't Hooft lines in 4d gauge theories. Upon reduction to three dimensions, the former give new scalar fields in 3d, while the latter can be seen as the 4d origin of monopole operators in 3d.

The structure of the VOA in the $\rQ_{1,2}$ cohomology of the 4d theory does not depend on $\beta$ as long as it is finite; $\beta$ simply determines the complex structure of the two-torus,
\begin{equation}
\tau = \frac{i\beta}{2\pi}.
\end{equation}
However, exactly at the $\beta=0$ point, the algebraic structure can jump: the torus degenerates to a circle, while the full 4d theory lands at the 3d fixed point. Importantly, precisely at this point, the space of local operators gets extended by wrapped lines, as explained around \eqref{ops_extension}, so the cohomology of local operators is expected to change as well. Since $\rQ_{1,2}$-closed operators remain closed under the map \eqref{ops_extension}, three things can happen:
\begin{enumerate}
	\item Closed and not exact operators from $\cV_{\rm 4d}^{\rm loc}$ can become exact in $\cV_{\rm 3d}^{\rm loc}$. We refer to this as lifting of the cohomology classes.
	\item The larger space $\cV_{\rm 3d}^{\rm loc}$ might contain additional closed and non-exact operators.
	\item Some closed and not exact operator from $\cV_{\rm 4d}^{\rm loc}$ might just vanish in $\cV_{\rm 3d}^{\rm loc}$.
\end{enumerate}
We assume that the third scenario only occurs when operators contain derivatives, at least for the Schur operators. In the chiral algebra, we only have holomorphic derivatives; in the 3d limit, they are equivalent to the anti-holomorphic derivatives, which are of course $\rQ_{1,2}$-exact. Thus, effectively, the third option in this list reduces to the first one. We will argue, and partly conjecture, that the first scenario indeed takes place, while the second does not, and furthermore, all the Higgs branch (\emph{i.e.} scalar Schur) operators, and only them, remain in the cohomology.

\subsection{Lifting of spinning Schur operators}

First, let us recall some basic facts. Quantum numbers of Schur operators in 4d --- the dimension $E_{\rm 4d}$, the $SU(2)_R$ charge $R$, the $U(1)_r$ charge $r$, and spins the $(j_1, j_2)$ --- obey the following \cite{Beem:2013sza}:
\begin{align}
E_{\rm 4d}-2R-j_1-j_2&=0,\cr
r + j_1 - j_2 &=0,
\end{align}
and conformal dimension of the corresponding chiral algebra operator is given by
\begin{equation}
h=R+j_1+j_2.
\end{equation}
The scalar Schur operators, \emph{i.e.}, those with $j_1=j_2=0$, are precisely the Higgs branch operators, which (together with a more general set of Hall-Littlewood operators) give primary operators in the VOA, which are also strong generators in the VOA.

Next in 3d $\cN=4$, where the R-symmetry algebra is $\mathfrak{su}(2)_H\oplus \mathfrak{su}(2)_C$, the analog of Schur operators are simply the Higgs branch operators. They are 3d scalars, $j=0$, have $\mathfrak{su}(2)_H$ charge $R$, have zero $\mathfrak{su}(2)_C$ charge, and obey \cite{Chester:2014mea,Beem:2016cbd}
\begin{equation}
E_{\rm 3d} = R.
\end{equation}

Now upon dimensional reduction, the Lorentz algebra $\mathfrak{su}(2)_M$ in 3d is identified as the diagonal subalgebra of the 4d Lorentz algebra $\mathfrak{su}(2)_\ell\oplus \mathfrak{su}(2)_r$. A local operator of 4d spins $(j_1, j_2)$ thus breaks into irreducible $\mathfrak{su}(2)_M$-representations according to the decomposition of $j_1\otimes j_2$. Only a very specific component in this decomposition is relevant for the Schur operators, however.  Recall that it was shown in \cite{Beem:2013sza} that the Schur operators in 4d are necessarily highest weight states for $\mathfrak{su}(2)_\ell\oplus \mathfrak{su}(2)_r$, \emph{i.e.}, components with the largest eigenvalues of $j_1$ and $j_2$. Under the diagonal subalgebra $\mathfrak{su}(2)_M$, the highest weight operator has the eigenvalue $j_1+j_2$, which is non-zero unless $j_1=j_2=0$. Because all operators in the 3d cohomology ought to be scalars, this immediately implies that spinning Schur operators, \emph{i.e.}, those with $(j_1, j_2)\neq (0,0)$, must be lifted in the 3d limit by some line operators. So this very simple representation-theoretic argument implies

\begin{proposition}
	All Schur operators of spins $(j_1, j_2)\neq (0,0)$ are lifted from the cohomology in the 3d limit.
\end{proposition}

\subsection{The fate of scalar Schur operators}
It remains to understand what happens to scalar Schur operators, that is the Higgs branch operators. The standard lore that the Higgs branch is unaffected by the KK reduction allows to make a natural guess: we expect all the 4d Higgs branch operators to survive the 3d limit (call it property A), and no new operators to be added (call this property B).

Unlike the Proposition 1 (which was very easy to prove), this one appears to be quite a non-trivial statement, so we will essentially resort to making a conjecture. Note that we can make the following partial argument in favor of property B. Suppose that we found a line operator $\cL$ that is closed under $\cQ^H=\rQ_1 + \rQ_2$, and we test the assumption whether it can give a new local cohomology class in 3d. Assume that it is invariant under translations along its worldline, so when we wrap it on $S^1$, we have $E\cL=0$. The $\cQ^H$-exact twisted-translation $\bar{L}_{-1}$ acts on it as $\bar{L}_{-1}\cL=(P_\varphi - 2R + E)\cL = (P_\varphi-2R)\cL$. Thus we say that $(P_\varphi-R)\cL = R\cL + \{\cQ^H,\dots\}$. In the 3d limit, $P_\varphi-R$ becomes the twisted translation along the preferred circle on $S^3$, which is also a $\cQ^H$-exact operation. Thus $R\cL$ must be $\cQ^H$-exact in 3d, which says that either $\cL$ is exact, or it has $R=0$. Since in the cohomology, the 3d conformal dimension $E_{\rm 3d}=R$, this implies $E_{\rm 3d}=0$, which can only be the case for the identity operator. This suggests that $\cL$ cannot produce a new local operator. Certainly it would be interesting to understand this, and especially the property A, better. For now, we will proceed with the

\begin{conjecture}
	The vector space of Higgs branch Schur operators is not affected by the KK reduction.
\end{conjecture}

\begin{remark}
	For the Coulomb branch supercharges $\cQ_{1,2}^C$ of \cite{Dedushenko:2017avn,Dedushenko:2018icp}, the opposite is true. In this case, the 4d cohomology contains line operators (Wilson, 't Hooft and dyon lines), but no local operators. Upon reduction to 3d, these lines reduce to local cohomology classes.
\end{remark} 

\begin{remark}
	For scalar Schur operators, $E_{\rm 4d}=2R$ and $E_{\rm 3d}=R$, which manifests a non-trivial renormalization: their conformal dimension changes along the flow.
\end{remark}

\subsection{Lifting in Lagrangian theories}
A very basic example of lifting by lines occurs in the dimensional reduction of Lagrangian 4d $\cN=2$ theories. In this case, free hypers and free vectors give $\beta\gamma$ and small $bc$ systems in the VOA, and for interacting theories, the vertex operators are constructed from $\beta, \gamma, b$ and $\partial c$ in a way dictated by the BRST cohomology \cite{Beem:2013sza}.

Upon dimensional reduction, Wilson lines wrapping the $S^1$ give rise to new operators in 3d, and an extra scalar in the 3d $\cN=4$ vector multiplet can be given such an interpretation. Denoting this extra scalar by $\phi$, a straightforward computation shows that
\begin{equation}
\rQ_1 \phi \propto b,\quad \rQ_2 \phi \propto \partial c,
\end{equation}
where we have identified $b$ and $\partial c$ as the appropriate gaugini components. This immediately implies that all 4d cohomology classes that involve small $bc$ ghosts in their expressions become Q-exact in 3d, and only gauge-invariant combinations of $\beta\gamma$ remain in the cohomology, which is exactly the expected operator spectrum in 3d \cite{Dedushenko:2016jxl}.

\subsection{Correlators and algebraic structures}
Now that we know the fate of Schur operators at the level of vector spaces, -- the subspace of scalar operators is isomorphic to the 3d cohomology, -- it remains to study the algebraic structure. Namely, we seek to determine how the VOA structure on the space of Schur operators in 4d determines the associative algebra structure in 3d.

The way to answer this question is by studying correlation functions: in 4d they are captured by the VOA correlators on the torus. In the limit $\tau\to 0$, the torus degenerates to a circle, and torus correlators, properly renormalized, reduce to the $S^1$ correlators, which are equivalent to an associative algebra equipped with the (twisted) trace map. This data captures the 1d TQFT sector in 3d SCFT, with some interesting details uncovered along the way. 

We will argue that the $\tau\to0$ limit of (properly renormalized) torus correlators gives a \emph{degenerate} trace map on the non-commutative Zhu algebra. Any correlator involving operators that are lifted from the cohomology in the 3d limit vanishes as $\tau\to0$. Such operators ``decouple'' from correlators and form an ideal inside the associative algebra. This property is a consequence of physics and should be regarded as a constrain on the VOAs that show up in the 4d SCFTs.

To properly identify the algebra in 3d, we quotient by this ideal, and this induces a \emph{non-degenerate} trace map on the quotient algebra. Even after doing this, however, we might end up with the deformation of the algebra one finds in the 3d SCFT.

In the next two sections, we proceed in two steps. First, we identify the (twisted) Zhu algebra as the dimensional reduction of a VOA. Then, we study the degeneration limit of torus correlators and argue that it determines a twisted trace on the Zhu algebra.

\section{Zhu algebra}\label{sec:Zhu}
\subsection{Dimensional reduction of a VOA}
Suppose we wanted to define the notion of dimensional reduction of a vertex algebra, independent of the present context. The natural way to do so would be to put our VOA on $\R\times S^1$, where $S^1$ has circumference $\beta$, with periodic boundary conditions. Then one takes the limit $\beta\to0$ and only keeps the zero modes of vertex operators.

A slightly more pedagogical way is to consider on $\R\times S^1$ some general insertions $\cO_1(w_1)\dots \cO_n(w_n)$ that are separated by distances of order $1$ in the $\R$ direction, perform a scale transformation by a factor of $1/\beta$ (after which the circle has circumference $1$, and the vertex operators are separated in the $\R$ direction by distances $\sim1/\beta$), and then take the limit $\beta\to0$. On the cylinder, $\cO(w) = e^{-iw T_0}\cO(0) e^{iwT_0}$, where $T_0\propto L_0 - c_{\rm 2d}/24$ is the holomorphic Hamiltonian, so in the infinitely long tubes between the operator insertions, the factors $e^{iwT_0}$ act as projectors $\Pi$ to the lowest-weight states on a circle (plus there is a constant phase describing Lorentz rotation). Hence the algebra in the $\beta\to0$ limit is determined by the algebra of zero modes, defined as $\Pi\cO(w)\Pi$ for each operator on the cylinder.

Suppose we have a VOA with conformal $\Z$-grading on $\C$ rather than the cylinder, and a vertex operator $\cO$ written in the math conventions,
\begin{equation}
\cO(z) = \sum_n \cO_n z^{-n-1}.
\end{equation}
Its zero mode is the component that has zero weight and thus preserves the lowest-weight subspace of any module. This component is $\cO_{\deg(\cO)-1}$, and the standard notation in the math literature is
\begin{equation}
o(\cO) \equiv \cO_{\deg(\cO)-1}.
\end{equation}
A primary operator on the cylinder and in the flat space are related by $\cO(z) = (iz/\ell)^{-\deg(\cO)}\cO(w)$, which implies the following relation between the two notions of zero modes:
\begin{equation}
\label{prime_zero}
o(\cO) = (i/\ell)^{-\deg(\cO)}\Pi\cO(w)\Pi.
\end{equation}
For descendant operators, additional terms appear in the conformal transformation, but after taking the necessary projectors, only zero modes survive, and the general form of the relation is:
\begin{equation}
\label{desc_zero}
\Pi\cO(w)\Pi = (i/\ell)^{\deg(\cO)}\left( o(\cO) + \sum_i o(\cO_i) \right),
\end{equation}
where $\cO_i$ are some other operators. Precisely what operators appear on the right does not matter at this point. What matters is that the algebra of either $\Pi\cO(w)\Pi$ or $o(\cO)$ zero modes (they are related by a change of basis \eqref{desc_zero}) provides the natural definition of dimensional reduction for the VOA. We will see later that the high-temperature limit of torus correlators is also controlled by this algebra. Such algebras in fact play central role in the theory of vertex algebras, and are known as the Zhu algebras, first introduced in \cite{Zhu}. It turns out that one can define a new product on the VOA V, denoted by $\star$,
\begin{equation}
\star: V\otimes V \to V,
\end{equation}
such that
\begin{equation}
o(\cO_1)o(\cO_2)\big|_{M^{(0)}} = o(\cO_1\star \cO_2)\big|_{M^{(0)}},
\end{equation}
where $\big|_{M^{(0)}}$ means that the operator acts on the lowest-weight space of some VOA module $M$. This product defines an associative algebra structure on a certain quotient of the VOA known as the Zhu algebra \cite{Zhu}, which we denote $\zhu(V)$ following \cite{2016arXiv160500138A}  (not to confuse with Zhu's $C_2$-algebra $R_V$, which is supercommutative and has previously appeared in the context of 4d SCFTs in \cite{Beem:2017ooy}).

In our present context, we have to work with more general $\frac12 \Z$-graded VOSAs, and as it turns out, we also have to consider their twisted modules. This requires a slightly more general notion of Zhu algebra, sometimes referred to as twisted Zhu algebra. For that reason, we now present a more precise (but still brief) overview of the necessary notions.

\subsection{Review of Zhu algebra}\label{sec:Zhu_math}
Let us now review the construction and main properties of the Zhu algebra. The original definition in \cite{Zhu} was for $\Z$-graded bosonic VOAs, but we need to work in a more general context of $\frac12\Z$-graded super-VOAs that do not satisfy the spin-statistics relation. Furthermore, the definition in \cite{Zhu} was relevant for the study of untwisted modules. In our case, we look at the VOA on $\R\times S^1$ with periodic boundary condition along $S^1$ (R sector), and upon the standard conformal map to $\C$, the Ramond puncture is generated at the origin: operators of half-integral spin are not single-valued. This does not happen for standard (NS sector) modules, in other words, we have to include twisted modules into consideration.

It is not hard to extract the necessary definitions and theorems from the literature: the case of twisted modules for ordinary VOAs was studied in \cite{Dong1998}; twisted modules for $\frac12\Z$-graded vertex operator superalgebras satisfying the spin-statistics relation were considered in \cite{Dong_Zhao_06}. As usual, all it takes to generalize to the super-case is replace commutators by graded-commutators, \emph{i.e.} add signs in various formulas (see e.g. \cite{Abe2007} for the definition of VOSA that does not assume spin-statistics). Because of that, the spin-statistics relation assumed in \cite{Dong_Zhao_06} does not really limit the applicability of their results, and we can readily extend them to our case of interest. Another important reference is \cite{Van_Ekeren_2013}. Let us review the basic definitions,

\begin{definition}
	A Vertex Operator Superalgebra (VOSA) is the data $(V, Y(\cdot, z), \mathbf{1}, \omega)$, where 
	\begin{equation}
	V=\oplus_{n\in\frac12\Z_{\geq 0}} V_n=V^{\underline{0}}\oplus V^{\underline{1}}
	\end{equation} 
	is a $\frac12 \Z$-graded super vector space, $Y(\cdot, z): V \to {\rm End} V[[z,z^{-1}]]$ is the vertex operator map, and $\mathbf{1}, \omega\in V^{\underline{0}}$ are the distinguished unit and Virasoro element respectively. These data satisfy the standard axioms, which we do not list for brevity (see \cite{kacvertex,Dong_Zhao_06}, we follow \cite{Dong_Zhao_06}), in particular the $\frac12\Z$-grading is the conformal grading by $L_0$. Notice that (unlike in \cite{Dong_Zhao_06}) we do not assume that the $\frac12 \Z_{\geq 0}$ grading $\oplus_{n\in\frac12\Z_{\geq 0}} V_n$ and the $\Z_2$-grading $V^{\underline{0}}\oplus V^{\underline{1}}$ are related (no spin-statistics relation). Let us use the math conventions for labeling modes,
	\begin{equation}
	Y(a,z) = \sum_{n\in\Z} \frac{a_n}{z^{n+1}}.
	\end{equation}
\end{definition}

\begin{definition}
	An automorphism of a VOSA $V$ is a linear automorphism of $V$ that preserves $\omega$ and is compatible with $Y$, \emph{i.e.} $g Y(a,z) g^{-1}=Y(ga, z)$.
\end{definition}

There exists a special automorphism ${s}\in{\rm Aut}(V)$ defined by ${s}\big|_{V_n} =(-1)^{2n}$, \emph{i.e.}, it multiplies elements of half-integer conformal weight by $-1$. This automorphism will play special role in the following. Let us also fix $g\in {\rm Aut}(V)$ of finite order $T_0$, and denote the order of $g{s}$ by $T$. Following \cite{Dong_Zhao_06}, we define weak $g$-twisted $V$-modules and admissible $g$-twisted $V$-modules (\cite{Dong_Zhao_06} also define ordinary $g$-twisted modules, but we do not need them here). 

\textbf{Physics digression. } \emph{It might be useful to give some intuition before diving into definitions. Operators in the VOSA are always mutually local, (up to signs originating from odd parity of fermions,) no matter what modules we study. When we consider usual (non-twisted) modules of a VOSA, we think of them as new operators inserted at the origin of $\C$: different such operators are not necessarily mutually local, but they are still local relative to the vertex operators belonging to the VOSA. The latter means that vertex operators from the VOSA are still single-valued on $\C$. Vertex operators act on the module via the OPE with this new operator at the origin. When we consider twisted modules, however, we also abandon locality of vertex operators relative to the new operator (corresponding to the twisted module) inserted at the origin. In this case a vertex operator is not single-valued and may acquire some phase when translated around the origin. Since we picked $g$ of finite order $T_0$, this phase must be a $T_0$-th root of unity. Now if we use the standard conformal map $z/\ell=e^{iw/\ell}$ to put VOSA on the cylinder $S^1\times \R$, the circle around the origin of $\C$ becomes the circle of $S^1\times\R$, and periodicity of observables changes upon this mapping. Operators of half-integer weights acquire an additional minus sign around the circle, so the new phase is dictates by $g{s}$ rather than $g$. Because $g{s}$ has order $T$, the operators on $S^1\times \R$ can acquire a phase given by the $T$-th root of unity when translated around the $S^1$. One important observation for the constructions discussed later is that only vertex operators fixed by $g{s}$ are periodic and thus have a zero mode on $S^1\times\R$.}

Consider decompositions of $V$ into eigenspaces with respect to the actions of $g{s}$ and $g$,
\begin{align}
V&=\oplus_{r\in\Z/T\Z} V^{*r},\cr
V&=\oplus_{r\in\Z/T_0\Z} V^r.
\end{align}
\begin{definition}
	A weak $g$-twisted $V$-module is a vector space $M$ equipped with the linear map
	\begin{align}
	V &\to ({\rm End})[[z^{1/T_0}, z^{-1/T_0}]]\cr
	a &\mapsto Y_M(a,z)=\sum_{n\in \frac1{T_0}\Z} a_n^M z^{-n-1}, \text{ where } a_n^M\in {\rm End} M,
	\end{align}
	such that for all $0\leq r\leq T_0-1$, $u\in V^r$, $v\in V$, $w\in M$,
	\begin{align}
	\label{module_ax}
	Y_M(u,z)&=\sum_{n\in \frac{r}{T_0}+\Z} u_n^M z^{-n-1},\cr
	u_n^M w &=0 \text{ for } n\gg 0,\cr
	Y_M(\mathbf{1},z)&={\rm Id}_M,\cr
	z_0^{-1}\delta\left( \frac{z_1-z_2}{z_0} \right)Y_M(u,z_1)Y_M(v,z_2) &- (-1)^{\widetilde{u}\widetilde{v}} z_0^{-1}\delta\left( \frac{z_2-z_1}{-z_0}\right)Y_M(v,z_2)Y_M(u,z_1)\cr
	=z_2^{-1}\left( \frac{z_1-z_0}{z_2} \right)^{-r/T_0}&\delta\left( \frac{z_1-z_0}{z_2} \right) Y_M(Y(u,z_0)v,z_2),
	\end{align}
	where $\widetilde{u}$ is the $\Z_2$ parity of $u$.
\end{definition}
The usual Virasoro axioms do not need to be included: they are part of the VOSA definition, and for the twisted module, their analogs follow from \eqref{module_ax}. The notion of admissible $g$-twisted $V$-modules is given by \cite{Dong_Zhao_06}:

\begin{definition}
	An admissible $g$-twisted $V$-module is a weak $g$-twisted $V$-module $M$ graded by $\frac{1}{T}\Z_{\geq 0}$,
	\begin{equation}
	M = \oplus_{n\in\frac1{T}\Z_{\geq 0}} M(n),
	\end{equation}
	such that for $a\in V$ homogeneous with respect to the $\frac12\Z_{\geq 0}$ degree $\deg(a)$, 
	\begin{equation}
	a_m^M M(n) \subseteq M(n+\deg(a)-m-1).
	\end{equation}
\end{definition} 
In \cite{Dong_Zhao_06}, an associative algebra for every $g\in{\rm Aut}(V)$ is constructed, which we denote here by $\zhu_g(V)$. This is the generalization of the famous Zhu algebra (constructed originally in \cite{Zhu} for non-twisted modules of VOAs,) to the case of $g$-twisted modules of VOSAs \cite{Dong_Zhao_06}. The main reason \cite{Zhu,Dong_Zhao_06} study this algebra is that it controls the representation theory of $V$. The ($g$-twisted) modules of $V$ give rise to the modules of the ``zero modes algebra'' $\zhu_g(V)$, and it turns out that the opposite is also true: given a module for $\zhu_g(V)$, the rich and rigid structure of vertex algebras guarantees that it can be uniquely (up to an isomorphism) extended to the $g$-twisted $V$-module. More precisely, \cite{Dong_Zhao_06} prove equivalence of the categories of completely reducible $\zhu_g(V)$-modules and completely reducible admissible $g$-twisted $V$-modules. We mentioned this important result for completeness of presentation, however it does not play any role in the current paper. What is interesting to us here is the algebra $\zhu_g(V)$ itself, so we give its definition and properties.

\begin{remark}
	$\zhu_g(V)$ will be filtered by the conformal degree \cite{2016arXiv160500138A}, and it is convenient to adjoin a formal parameter $\hbar$ of degree $\deg(\hbar)=1$ that compensates for the mismatch of degree in products and makes it into a graded algebra. It makes arguments more intuitive, and passing to the associated graded simply corresponds to the $\hbar\to 0$ limit. One can in fact develop a theory of $\hbar$-deformed vertex operators, see \cite{DSK}, but we will not need it in full generality here. We will simply reintroduce $\hbar$ when borrowing results from the math literature. Later, we will identify $\hbar$ as a number proportional to $1/\ell$. Following \cite{DSK}, one could use the notation $\zhu_{\hbar, g}(V)$ when $\hbar$ is included, but we will stick to $\zhu_g(V)$, as the meaning should be clear from the context.
\end{remark}

Let $v\in V$ and $u\in V^{*r}$, and let us set $\delta_r=1$ if $r=0$ and $\delta_r=0$ otherwise. Define two new products on $V$:
\begin{align}
\label{products}
u\circ_g v &= {\rm Res}_z \frac{(1+\hbar z)^{\deg(u) - 1 + \delta_r +\frac{r}{T}}}{z^{1+\delta_r}} Y(u,z)v,\cr
u\star_g v&= \begin{cases}
{\rm Res}_z Y(u,z)\frac{(1+\hbar z)^{\deg(u)}}{z}v &\text{ if } r=0,\\
0 &\text{ if } r>0,
\end{cases}
\end{align}
a vector subspace
\begin{equation}
O_g(V) = V\circ_g V,
\end{equation}
and a quotient vector space
\begin{equation}
\zhu_g(V) = V/O_g(V).
\end{equation}
Also define the ``zero mode on the cylinder'' operation on $V$, which was motivated previously (see discussion on why we have to look at the subspace fixed by $g{s}$). For homogeneous $a\in V^{*r}$, and a weak $g$-twisted $V$-module $M$, define 
\begin{equation}
o^M_g(a) = \begin{cases}
a^M_{\deg(a)-1}, &\text{if } a\in V^{*0}, \text{ \emph{i.e.} } g{s}(a)=a,\\
0, &\text{otherwise}.
\end{cases}
\end{equation}
(It is easy to see that $\deg(a)-1$ always belongs to the correct set of indices.)

Now we list some properties of the objects defined so far (see \cite{Zhu}, \cite{Dong_Zhao_06} and \cite{DSK} for the proofs):
\begin{enumerate}
	\item If $r\neq 0$, then $V^{*r}\subseteq O_g(V)$, so $\zhu_g(V) = V^{*0}/(O_g(V)\cap V^{*0})$. This also shows that $\zhu_g(V)$ is a quotient of $\zhu_{g={\rm id}}(V^{*0})$.
	\item $O_g(V)$ is a two-sided ideal of $V$ under $\star_g$.
	\item Product $\star_g$ induces an associative algebra structure on $\zhu_g(V)$.
	\item The class $[\mathbf{1}]\in\zhu_g(V)$ of $\mathbf{1}\in V$ is the unit of $\zhu_g(V)$.
	\item The class $[\omega]\in \zhu_g(V)$ of the Virasoro element lies in the center of $\zhu_g(V)$.
	\item $L_{-1} a + \hbar L_0 a \in O_g(V)$ for any $a\in V$.
	\item For any weak $g$-twisted $V$-module $M$, and any $a,b\in V$, we have 
	\begin{align}
	o_g^M(a\circ_g b)\big|_{\hbar=1}&=0,\\
	o_g^M(a)o_g^M(b)\big|_{\hbar=1}&= o_g^M(a\star_g b)\big|_{\hbar=1},
	\end{align}
	when acting on the lowest weight subspace of $M$.
\end{enumerate}
These properties clearly demonstrate that $\zhu_g(V)$ is the correct ``algebra of zero modes'' formalizing the notion of dimensional reduction of a vertex algebra. In the next section we will explain its relevance for the high-temperature limit of torus correlators. In this subsection, it only remains to specify $g\in{\rm Aut}(V)$. Recall the basic notions of sectors:

\begin{definition}
	The Neveu-Schwarz (NS) sector weak $V$-modules are those with $g={\rm id}$, \emph{i.e.}, the usual untwisted weak $V$-modules. The Ramond (R) sector weak $V$-modules are the weak $g$-twisted $V$-modules with $g={s}$. Similarly for admissible $V$-modules.
\end{definition}

Note that for $g={s}$, $g{s}={\rm id}$, so $V^{*0}=V$, \emph{i.e.}, all vertex operators are periodic on the cylinder, which is of course standard for the R sector. We know that the corresponding zero modes algebra is what we are after, and from this subsection we conclude that this algebra is $\zhu_{s}(V)$.\footnote{This $\zhu_{s}(V)$ is the same as $\zhu_H V$, or rather $\zhu_{\hbar,H} V$, that appears in \cite{DSK}.}

\textbf{Definition/Proposition:}\footnote{This is a definition in the math sense, and a proposition in the physics sense, where we already have an intuitive notion of what the dimensional reduction means.} dimensional reduction of a $\frac12\Z_{\geq 0}$-graded VOSA is the associative filtered algebra $\zhu_{s}(V)$. 

We could also study a twisted dimensional reduction, with operators periodic around the $S^1$ only up to an automorphism $g$. In such a case, the corresponding algebra would of course be $\zhu_{g{s}}(V)$. In this note, we limit ourselves to the ordinary, untwisted dimensional reduction.

\section{Torus correlators and the trace}\label{sec:trace}

\subsection{Torus correlators}
For every (possibly twisted) module $M$ of the VOA, one can introduce the notion of a torus correlation function given by the character of this module modified by additional insertions of vertex operators under the trace \cite{Zhu}, 
\begin{equation}
\label{tor_corr0}
\trace_M\left\{ (-1)^F q^{L_0 - c_{\rm 2d}/24}\cO_1(w_1)\dots \cO_n(w_n)\right\}.
\end{equation}
In the $\frac12\Z$-graded case, since there are four Spin structures on the torus, there are four variants of such characters and correlators. They come from the choice of module (the NS or the R sector module), and from the additional twist by ${s}\in{\rm Aut}(V)$ in the trace.\footnote{Recall that ${s}$ multiplies half-integrally graded operators by $-1$.} Quite generally, if there is an automorphism $g\in{\rm Aut}(V)$, one can define a different module structure on $M$, denoted $g\cdot M$, by $Y^{g\cdot M}(a,z) = Y^M(g(a),z)$. If module $g\cdot M$ is isomorphic to $M$, we say that it is $g$-invariant, and there exists $g_M \in {\rm End}(M)$, such that $Y^M(g(a),z) = g_M Y^M(a,z) g_M^{-1}$. In this case, one can insert this $g_M$ under the trace over $M$ to define a ``$g$-twisted'' trace ${\rm Tr}_M \{g_M \dots\}$. In particular for us, the untwisted trace (that can also be referred to as the R trace because it corresponds to periodic boundary conditions along the thermal circle) gives \eqref{tor_corr0}, while the ${s}$-twisted trace  (or the NS trace) can be used to define twisted torus correlators over some ${s}$-invariant module $M$,
\begin{equation}
\trace_M\left\{ (-1)^F {s}_M q^{L_0 - c_{\rm 2d}/24}\cO_1(w_1)\dots \cO_n(w_n)\right\}.
\end{equation}

Recall that we are mostly interested in the NS-R sector correlators, more specifically vacuum torus correlators defined using the untwisted trace. As a module $M$ we choose the NS sector \emph{vacuum} module $V$ (the Fock space underlying the VOA). This corresponds to the 4d theory on $S^3\times S^1$ without any extra defects or holonomies. The normalized torus correlators in this case are
\begin{equation}
\label{tor_corr}
\langle \cO_1(w_1) \dots \cO_n(w_n)\rangle = \frac{\trace_V (-1)^F q^{L_0 - c_{\rm 2d}/24}\cO_1(w_1)\dots \cO_n(w_n)}{\trace_V (-1)^F q^{L_0 -c_{\rm 2d}/24}},
\end{equation}
where the operators are properly ordered, $y_1>\dots>y_n$, for $y=-\Im(w)$, and the nome is
\begin{equation}
q=e^{2\pi i\tau},\quad \tau=\frac{i\beta}{2\pi}.
\end{equation}
As we take the $\tau\to0$ limit, the Ramond Spin structure circle shrinks to zero size, the NS Spin structure circle remains finite, and the nome becomes $1$,
\begin{equation}
q\to 1.
\end{equation}

\subsection{High temperature limit and modularity}\label{sec:modularity}
As usual, it is hard to analyze correlators in the high temperature (\emph{i.e.} $\tau\to 0$) limit because the trace in \eqref{tor_corr} is not convergent at $q=1$. It is natural to perform a modular transformation,
\begin{align}
\label{modular_tr}
\widetilde\tau &= -\frac1{\tau},\cr
\widetilde{w} &= \frac{w}{\tau},
\end{align}
since in the dual variable, the limit $\widetilde\tau\to +i\infty$ is much more tractable. In the $\Z$-graded case, the behavior of torus correlators under this transformation is expected to be as follows:
\begin{equation}
\label{modular_corr}
\trace_{M_i}\left\{ (-1)^F q^{L_0 - c_{\rm 2d}/24}\cO_1(w_1)\dots \cO_n(w_n)\right\}=\tau^{-\sum_k h_k}\sum_j S_{ij} \trace_{M_j}\left\{(-1)^F \widetilde{q}^{L_0 - c_{\rm 2d}/24}\cO_1(\widetilde{w}_1)\dots \cO_n(\widetilde{w}_n)\right\},
\end{equation}
where $h_i$ is the conformal dimension of $\cO_i$, the sum goes over an appropriate class of irreducible modules $M_i$, and $S_{ij}$ is the modular S matrix. This expectation, though, is quite naive, and even making sense of the right hand side could be problematic, as we will explain later in this section.

In the more general $\frac12 \Z$-graded case of our interest, the NS-R spin structure switches into the R-NS under the S transform. This means that the NS sector character should transform into the sum of R sector twisted characters. The latter are given by the ${s}$-twisted traces over the R sector modules, and in order for such traces to make sense, the corresponding modules must be ${s}$-invariant. It was shown in \cite{Van_Ekeren_2013} that at least in the $C_2$-cofinite case, only ${s}$-invariant modules appear via the modular transformation. We are going to conjecture that the same holds for our class of non-$C_2$-cofinite VOAs, so we can pick ${s}_M\in {\rm End}(M)$ such that ${s}(a) = {s}_M a {s}_M^{-1}$ when acting on this module.\footnote{At least $M\oplus {s}\cdot M$ is always ${s}$-invariant, even if $M$ itself is not. However, we are not aware of any examples where characters of such reducible modules must be included for modularity.}  Hence the formula \eqref{modular_corr} generalizes to
\begin{align}
\label{modular_corr_mod}
&\trace_{M_i}\left\{ (-1)^F q^{L_0 - c_{\rm 2d}/24}\cO_1(w_1)\dots \cO_n(w_n)\right\}\cr
&=\tau^{-\sum_k h_k}\sum_j S_{ij} \trace_{\widetilde{M}_j}\left\{{s}_{\widetilde{M}_j}(-1)^{F} \widetilde{q}^{L_0 - c_{\rm 2d}/24}\cO_1(\widetilde{w}_1)\dots \cO_n(\widetilde{w}_n)\right\},
\end{align}
where now $M_i$ is the NS-sector irreducible module, and the sum is over the R-sector ${s}$-invariant modules $\widetilde{M}_j$.

The status of this equation, and even of equation \eqref{modular_corr}, is far from being a theorem. Starting with the pioneering work of Zhu \cite{Zhu}, there have been a number of articles proving it for $C_2$-cofinite VOAs using various methods, considering both insertions of vertex operators and of more general intertwiners \cite{Dong:1997ea,2000math.....10180M,Huang:2003cq,miyamoto2004,Fiordalisi}, and including the case of twisted modules into the story \cite{Van_Ekeren_2013}. Yet, once we move outside the $C_2$-cofinite world, the knowledge is very limited, see however \cite{Creutzig:2012sd,Creutzig:2013yca,Arakawa:2016hkg,Auger:2019gts}.

In particular, the results of \cite{Creutzig:2013yca,Arakawa:2016hkg,Auger:2019gts} suggest that VOAs appearing in 4d SCFTs might still satisfy some version of the modularity properties \eqref{modular_corr},\eqref{modular_corr_mod}. Additionally, the work \cite{Beem:2017ooy} assumes this as well (at least the more conservative modularity of a character, thought of as a solution of the linear modular differential equation; but the $h_{\rm min}$ appearing there clearly suggests an interpretation as a conformal dimension of an actual VOA module, not just a power in the leading coefficient of the character).

Certainly, there is a physical expectation at least if the VOA under consideration can additionally arise as (a chiral algebra of) some 2d CFT. If such a CFT exists, then \eqref{modular_tr} can be understood as a diffeomorphism from the torus with complex structure $\tau$ to another torus with complex structure $\widetilde\tau$. Local operators see it as a rescaling of $w$ by a complex constant $\tau$ (composition of a dilatation with a Lorentz transformation), so chiral observables transform simply according to
\begin{equation}
\cO(w) = \tau^{-h}\cO(\widetilde{w}).
\end{equation}
The Hilbert space of a 2d CFT decomposes into a sum of modules for the VOA, and the requirement of modular invariance of the torus partition function suggests that these modules must simply reshuffle into each other under the modular transformation, implying \eqref{modular_corr}.

Even if the VOA appears as a chiral algebra of some 2d CFT that is not invariant, but rather transforms in a known way under modular transformations, we can still use it to infer \eqref{modular_corr}. In particular, in Lagrangian 4d $\cN=2$ theories, the VOA can be described as the gauged beta-gamma system, which perhaps can be used to prove \eqref{modular_corr} in those cases.

\begin{remark}\label{modularity_remark}
	One important problem with the equations \eqref{modular_corr}, \eqref{modular_corr_mod} is that the modules $M_j$ or $\widetilde{M}_j$ appearing on the right might have infinite-dimensional conformal weight eigenspaces. This makes the trace problematic, thus the right hand side potentially ill-defined, and raises a question of whether the objects we study even make sense. This question seems even more urgent than the modularity property itself. However, it is still possible that \emph{twisted} traces make sense. When the VOA has enough automorphisms (say, originating from flavor symmetries), one can introduce generic twist parameters corresponding to those symmetries, which regularize twisted traces, and result in finite answers. Each individual twisted trace over $M_j$ would diverge when we remove the regularization, but the precise linear combination appearing on the right of \eqref{modular_corr} or \eqref{modular_corr_mod}, divided by the empty torus partition function (to define normalized correlators), might still be finite in this limit. Furthermore, as we compute normalized correlators, the divergence due to Cardy behavior should cancel out too. The regularization via twisted traces is familiar from the study of associative algebras (1d protected sectors) in 3d $\cN=4$ theories: it was argued in \cite{Gaiotto:2019mmf} that the twisted trace can be decomposed as a linear combination of twisted traces over the Verma modules of the algebra (corresponding to massive vacua). Each twisted trace over the individual Verma module is divergent in the no-twist limit, but the precise linear combination of traces picked out by the physical theory remains finite. Furthermore, in the context of VOA, all the applications we consider in the related work \cite{DW} show that the assumption of modularity gives correct results (at least when $c_{\rm 4d}>a_{\rm 4d}$).
\end{remark}

So we are going to assume \eqref{modular_corr} (or \eqref{modular_corr_mod} when appropriate) in this sense, and use it to describe the $\tau\to +0i$ limit of torus correlators. In the dual variable, this limit becomes $\widetilde{q}\to0$, so the factor of $\widetilde{q}^{L_0-c_{\rm 2d}/24}$ can be seen as a projector on the lowest-$L_0$ subspace. More precisely, suppose that the vertex operators sit at distinct locations along the $\varphi$ circle,
\begin{equation}
0<\varphi_1 < \varphi_2 <\dots <\varphi_n<2\pi,\quad \text{where } \varphi_i=\frac1{\ell}\Re(w_i).
\end{equation}
Then after the transformation $\widetilde{w}=w/\tau$, because we take $\tau=i\beta/(2\pi)$, we have
\begin{equation}
\Im(\widetilde{w}_i)=-\frac{2\pi\ell}{\beta}\varphi_i,
\end{equation}
and so they will sit at the distinct locations along the $\Im(\widetilde{w})$ direction,
\begin{equation}
-\frac{4\pi^2\ell}{\beta}<\Im(\widetilde{w}_n)<\dots<\Im(\widetilde{w}_2)<\Im(\widetilde{w}_1)<0.
\end{equation}
Recall also that on the cylinder, for chiral vertex operators, one has:
\begin{equation}
\hat\cO(\widetilde{w}) = e^{-i\widetilde{w}T_0}\hat\cO e^{i\widetilde{w}T_0},
\end{equation}
where $\hat\cO=\cO(0)$, and $T_0$ is the holomorphic Hamiltonian on the cylinder,
\begin{equation}
T_0=\frac1{\ell}\left(L_0 - \frac{c_{\rm 2d}}{24}\right).
\end{equation}
We use it to write the correlation function \eqref{modular_corr_mod} in the Hamiltonian-like form as
\begin{align}
\label{Ham_repr}
\tau^{-\sum_k h_k}\sum_{j} S_{ij}\trace_{\widetilde{M}_j}\left\{{s}_{\widetilde{M}_j} (-1)^{F}  e^{-i\widetilde{w}_1T_0}\widehat\cO_1 e^{i(\widetilde{w}_1-\widetilde{w}_2)T_0} \widehat\cO_2 e^{i(\widetilde{w}_2-\widetilde{w}_3)T_0} \dots e^{i(\widetilde{w}_{n-1}-\widetilde{w}_n)T_0} \widehat\cO_ne^{i\widetilde{w}_n T_0}\,\widetilde{q}^{T_0} \right\}.
\end{align}
This expression makes the $\beta\to 0$ limit transparent. Indeed, by writing
\begin{equation}
i(\widetilde{w}_k - \widetilde{w}_{k+1}) = \frac{2\pi\ell(\varphi_k - \varphi_{k+1})}{\beta} - \frac{2\pi i(y_k-y_{k+1})}{\beta}
\end{equation}
and recalling that $y \in (0,\beta\ell)$, we see that the imaginary part of $i(\widetilde{w}_k - \widetilde{w}_{k+1})$ remains finite, while the real part goes to $-\infty$. Hence $e^{i(\widetilde{w}_k-\widetilde{w}_{k+1})T_0}$ for $\beta\to 0$ acts as a projector on the subspace of lowest $T_0$ eigenvalue. Since each lowest-weight module $\widetilde{M}_j$ (of weight $\widetilde\Delta_j$) is graded by dimension,
\begin{equation}
\widetilde{M}_j = \oplus_{n \in \frac12 \Z_{\geq 0}} \widetilde{M}_j^{(n)},
\end{equation}
where the conformal dimension of $\widetilde{M}_j^{(n)}$ is $\widetilde\Delta_j + n$, we see that $e^{i(\widetilde{w}_k-\widetilde{w}_{k+1})T_0}$ projects on the lowest-weight subspace $\widetilde{M}_j^{(0)}$, and there is a remaining phase factor $e^{- \frac{2\pi i(y_k-y_{k+1})}{\beta\ell}\widetilde\Delta_j}$ from the imaginary part of $i(\widetilde{w}_k - \widetilde{w}_{k+1})$ that stays finite as we take $\beta\to0$. Finally, in the sum \eqref{Ham_repr}, by analogy with \cite{Beem:2017ooy}, the leading contribution comes from the modules $\widetilde{M}_{\rm min}$ of minimal conformal dimension $\widetilde\Delta_j = \widetilde\Delta_{\rm min}$. We denote the projector on the lowest-weight subspace as
\begin{equation}
\Pi: \widetilde{M}_j \to \widetilde{M}_j^{(0)},
\end{equation}
so the leading behavior of \eqref{Ham_repr} can be written as
\begin{equation}
\tau^{-\sum_k h_k} S_{ij}\widetilde{q}^{\widetilde\Delta_{\rm min}-c_{\rm 2d}/24}\trace_{\widetilde{M}_{\rm min}^{(0)}}\left\{{s}_{\widetilde{M}_j}(-1)^F  \hat\cO_1 \Pi \hat\cO_2 \Pi \dots \Pi \hat\cO_n\right\},
\end{equation}
where for simplicity we assumed that there exists only one module $\widetilde{M}_j$ of lowest weight $\widetilde\Delta_{\rm min}$, and the generalization to multiple such modules is obvious. We kept $F$ here that denotes the fermion number of $|v_0\rangle$ for generality. Finally, this allows to find the $\tau\to0$ behavior of normalized correlators \eqref{tor_corr} on the torus:
\begin{equation}
\langle \cO_1(w_1)\dots \cO_n(w_n)\rangle \sim \tau^{-\sum_k h_k} \frac{\trace_{\widetilde{M}^{(0)}_{\rm min}}\left\{{s}_{\widetilde{M}_{\rm min}^{(0)}}(-1)^F \hat\cO_1 \Pi \hat\cO_2 \Pi \dots \Pi \hat\cO_n\right\} }{\trace_{\widetilde{M}^{(0)}_{\rm min}}\left\{{s}_{\widetilde{M}_{\rm min}^{(0)}}(-1)^F\right\}}.
\end{equation}
If we define renormalized operators by
\begin{equation}
\label{renorm}
\cO_k^{\rm r}(w_k) = (-i \tau)^{h_k}\cO_k(w_k),
\end{equation}
then the high-temperature behavior can be written as
\begin{equation}
\label{highT_corr}
\lim_{\tau\to0} \langle \cO^{\rm r}_1(w_1)\dots \cO^{\rm r}_n(w_n)\rangle = i^{-\sum_k h_k}  \frac{\trace_{\widetilde{M}^{(0)}_{\rm min}}\left\{{s}_{\widetilde{M}_{\rm min}^{(0)}}(-1)^F \hat\cO_1 \Pi \hat\cO_2 \Pi \dots \Pi \hat\cO_n\right\} }{\trace_{\widetilde{M}^{(0)}_{\rm min}}\left\{{s}_{\widetilde{M}_{\rm min}^{(0)}}(-1)^F\right\}}.
\end{equation}

\begin{remark}
	For unitary $\Z$-graded VOAs, the minimal conformal dimension is $\Delta_{\rm min}=0$, and the corresponding module is the vacuum module. In this case, the lowest weight subspace is spanned by the vacuum $|0\rangle$ (or the R-sector vacua $|0\rangle_\alpha$ in the $\frac12 \Z$-graded case), and the above equation simplifies drastically,
	\begin{equation}
	\lim_{\tau\to0} \langle \cO^{\rm r}_1(w_1)\dots \cO^{\rm r}_n(w_n)\rangle =i^{-\sum_k h_k}  \frac{\langle 0| \hat\cO_1 \Pi \hat\cO_2 \Pi \dots \Pi \hat\cO_n|0\rangle}{\langle 0|0\rangle}.
	\end{equation}
	For general non-unitary VOAs, which are of our primary interest, the situation is much more involved: they might have a non-trivial $\widetilde\Delta_{\rm min}$, and the corresponding module might even have infinite-dimensional weight eigenspaces, as we already mentioned earlier.
\end{remark}

\begin{remark}
	The ``renormalization'' ${\rm r}$ is defined in \eqref{renorm} for operators of definite conformal weights $h_k$, but can be extended to the whole $V$, ${\rm r}: V\to V$, by linearity.
\end{remark}

\begin{remark}
	$\tau\to0$ limits of characters, among other things, are related (for rational VOAs) to quantum dimensions via $\frac{S_{i0}}{S_{00}}=\lim_{\tau\to+ 0i}\frac{{\rm ch}[M_i](\tau)}{{\rm ch}[V](\tau)}$, see \cite{Dijkgraaf:1988tf}. For such quantum, or asymptotic, dimensions outside the $C_2$-cofinite case, see \cite{Creutzig:2013zza,Creutzig:2014nua,Creutzig:2016htk,Creutzig:2016uqu} and references therein.
\end{remark}

\subsection{The twisted trace on $\zhu_{s}(V)$}
In the previous subsection, the main conclusion was equation \eqref{highT_corr} describing the high-temperature limit of torus correlators. If operators $\cO_i(w)$ are primary, then by \eqref{prime_zero} we can replace $\Pi\hat\cO \Pi$ by $(i/\ell)^{\deg(\cO)} o(\cO)$ and continue \eqref{highT_corr} with the following chain of equalities:
\begin{align}
\label{trace_on_zhu}
\lim_{\tau\to 0} \langle \cO_1^{\rm r}(w_1)\dots \cO_n^{\rm r}(w_n)\rangle &=\left(\frac1{\ell}\right)^{\sum_k h_k}\frac{\trace_{\widetilde{M}^{(0)}_{\rm min}}\left\{{s}_{\widetilde{M}_{\rm min}^{(0)}}(-1)^F o_{s}^{\widetilde{M}_{\rm min}}(\cO_1)\dots o_{s}^{\widetilde{M}_{\rm min}}(\cO_n)\right\}}{\trace_{\widetilde{M}^{(0)}_{\rm min}}\left\{{s}_{\widetilde{M}_{\rm min}^{(0)}}(-1)^F\right\}}\cr
&=\left(\frac1{\ell}\right)^{\sum_k h_k}\frac{\trace_{\widetilde{M}_{\rm min}^{(0)}}\left\{{s}_{\widetilde{M}_{\rm min}^{(0)}}(-1)^F o_{s}^{\widetilde{M}_{\rm min}}(\cO_1\star\dots \star\cO_n)\right\}}{\trace_{\widetilde{M}^{(0)}_{\rm min}}\left\{{s}_{\widetilde{M}_{\rm min}^{(0)}}(-1)^F\right\}}\Bigg|_{\hbar=1}.
\end{align}
In the last equality we employed property 7 of the Zhu algebra mentioned previously. 

If operators are not primary, then a more complicated relation \eqref{desc_zero} must be used, yet the general structure remains the same. Namely, the $\tau\to0$ limit of torus correlators is captured by two pieces of data: the Zhu algebra at $\hbar=\frac1{\ell}$, and the map $T_{s}$ describing the $\tau\to0$ limit of the torus one-point function, defined as follows for homogeneous elements:
\begin{align}
\label{twtr_formula}
[\cO] &\mapsto \hbar^{\deg(\cO)} \times\left[  \frac{\trace_{\widetilde{M}_{\rm min}^{(0)}}\left\{{s}_{\widetilde{M}_{\rm min}^{(0)}}(-1)^F o_{s}^{\widetilde{M}_{\rm min}}(\cO)\right\}}{\trace_{\widetilde{M}^{(0)}_{\rm min}}\left\{{s}_{\widetilde{M}_{\rm min}^{(0)}} (-1)^F\right\}}\Big|_{\hbar=1}\right],\quad \text{where } \hbar = \frac1{\ell},
\end{align}
and extended to $T_{s}: \zhu_{s}(V)\big|_{\hbar=1/\ell} \to \C$ by linearity. 

\begin{remark}
	Because $o_g^M\left(O_g(V)\right)\big|_{\hbar=1}$ acts trivially on the lowest weight subspace of module $M$, and the class $[\cO]$ is defined modulo $O_g(V)$, the above map $T_{s}:\zhu_{s}(V)\big|_{\hbar=1/\ell}\to\C$ is well-defined. This map is the twisted trace.
\end{remark}

\begin{definition}[\cite{ERS1}]
	Given an associative filtered algebra $\cA$ and a filtration preserving $\psi\in{\rm Aut}(\cA)$ , a $\psi$-twisted trace is a linear map $T_\psi:\cA\to \C$, such that $T_\psi(\alpha\beta)=T_\psi(\beta\psi(\alpha))$, $\forall \alpha,\beta\in\cA$. This $T_\psi$ is said to be non-degenerate if $B_T(\alpha,\beta)=T_\psi(\alpha\beta)$ is non-degenerate, and $T_\psi$ is called strongly non-degenerate if $B_T$ is non-degenerate in each filtration degree.\footnote{Strong non-degeneracy enters the theorem on short star-products in \cite{ERS1}, and is simply called non-degeneracy there. The ``strong'' terminology was suggested to the author by P.Etingof.}
\end{definition}

In the R-NS Spin structure, the vertex operators of half-integer dimension are anti-periodic along the $\varphi$ direction. This is manifested by the presence of ${s}_{\widetilde{M}_{\rm min}^{(0)}}$ in \eqref{twtr_formula} and readily implies:
\begin{equation}
T_{s}(\cO_1\star \cO_2) = T_{s}(\cO_2\star {s}(\cO_1) ).
\end{equation}

\begin{definition}
	Equation \eqref{twtr_formula} defines an ${s}$-twisted trace on $\zhu_{s}(V)\big|_{\hbar=1/\ell}$. We can also choose to treat $\hbar=\frac1{\ell}$ as a formal variable again, define $T_s(\hbar)=\hbar$, and extended it to $T_{{s}}: \zhu_{s}(V) \to \C[\hbar]$ by linearity.
\end{definition}

However, it is not non-degenerate.

\begin{remark}
	For any $\psi$-twisted trace $T$, the form $B_T$ is symmetric up to the automorphism $\psi$, so the left and right kernels of $B_T$ coincide.
\end{remark}
\begin{definition}
	The null subspace $N\subset \cA$ for a $\psi$-twisted trace $T$ is the kernel of $B_T$.
\end{definition}

\textbf{More physics input: } \emph{In the context of VOSAs arising from the 4d $\cN=2$ SCFTs, the twisted trace $T_{s}$ always has a non-empty null subspace. Indeed, all the non-Higgs (\emph{i.e.}, spinning) Schur operators get lifted from the cohomology in the 3d limit. This implies that the torus correlators that contain the corresponding vertex operators must vanish in the $\tau\to 0$ limit. We refer to this by saying that the spinning Schur operators decouple at $\tau\to0$. This property should be regarded as a physical constraint on the VOSAs that can appear in this way.}

\emph{More precisely, the operators that decouple might not be exactly the spinning Schur operators because of the operator mixing phenomena that take place on $S^3\times S^1$ (due to the non-zero curvature of $S^3$). When operator $\cO$ gets lifted from the cohomology, it becomes redundant in the sense that it is cohomologous to another operator. But that other operator need not be zero, -- instead it can be a linear combination of lower-dimension operators (with difference in dimension compensated by integer powers of $\hbar$). So the most general decoupling pattern is}
\begin{equation}
\label{mixing}
\cO - \sum_{k\geq 1} \hbar^k \cO_k = \{\rQ_i,\dots\} \text{ in the 3d limit},
\end{equation}
\emph{where $\cO_k$ are some lower-dimension Schur operators. Without loss of generality, we can assume that $\cO_k$ are those that survive the 3d limit, \emph{i.e.} the scalar Schur operators (everything else sits in the Q-exact term).}

\emph{So, the image in $\zhu_{s}(V)$ of all the vectors from $V$ of the form \eqref{mixing} must lie inside the null subspace $N\subset \zhu_{s}(V)$. Furthermore, vertex operators corresponding to scalar Schur operators ought to give non-zero classes in $\zhu_{s}(V)/N$. This is because they are not lifted from the cohomology, and again physics implies another constraint: the torus two-point function in the $\tau\to0$ limit should give a \emph{strongly} non-degenerate bilinear form for such operators. Hence scalar Schur operators cannot land in the null subspace.}

\begin{remark}
	To summarize, there is a null space $N\subset\zhu_{s}(V)$ for $T_{s}$ such that $\zhu_{s}(V)/N$ is precisely the space of scalar Schur operators in 4d. This $N$ is always non-empty, because any VOSA that we study has a stress-energy tensor, which comes from a spinning Schur operator, gives a non-trivial element in $\zhu_{s}(V)$, and thus, up to $O(\hbar)$ terms, must be in $N$.
\end{remark}

It is clear that $N$ is in fact an ideal of $\zhu_{s}(V)$. We therefore can take the quotient, and $T_{s}$ will induce a non-degenerate twisted trace on this quotient:
\begin{definition}
	Define the algebra
	\begin{equation}
	\cA_H = \zhu_{s}(V)/N,
	\end{equation}
	with the non-degenerate twisted trace map induced by $T_{s}$, which we also call $T_{s}$.
\end{definition}
All our explanations imply the following
\begin{proposition}
	The $(\cA_H, T_{s})$ data determines the protected 1d sector in the cohomology of the 3d $\cN=4$ theory on $S^3$ that arises in the $\beta\to0$ limit.
\end{proposition}
\begin{remark} $T_s$ on $\cA_H$ is non-degenerate by construction, but not necessarily \emph{strongly} non-degenerate. Physics predicts that it must be strongly non-degenerate, which we thus assume.\footnote{This is because the CFT two-point function can be diagonalized, and operators do not mix with higher-dimension operators due to locality.}
	\end{remark}
\begin{remark}
	\cite{ERS1} prove that the data $(\cA_H, T_{s})$ determines a short star product, \emph{i.e.} the one satisfying the truncation condition from \cite{Beem:2016cbd}: for $\deg(v)=|v|\in\frac12\Z$, $a\star b$ truncates after the $O(\hbar^{2\min(|a|,|b|)})$ term. More precisely, \cite{ERS1} prove that there is a bijection between non-degenerate short star products and strongly non-degenerate $\psi$-twisted traces. Another important property \cite{Beem:2016cbd,ERS1} is evenness of the star product, namely that the $O(\hbar^{2n})$ terms are symmetric and the $O(\hbar^{2n+1})$ terms are anti-symmetric. Star products are even at the SCFT point, but for non-conformal deformations, evenness might be lost (see, \emph{e.g.}, examples in \cite{Dedushenko:2016jxl}). For those 4d theories that land exactly at the SCFT point in 3d, the star product following from the construction of this paper must be even. It would be interesting to determine a criterion for the VOA to give the even star product.
\end{remark}

\section{Further comments}\label{sec:comments}
\subsection{Affine VOA}
For illustration purposes, let us briefly consider the case of an affine VOA $V_k(\mathfrak{g})$ at a non-critical level, leaving further applications to \cite{DW}. The basic OPE is given by
\begin{equation}
J_A(z)J_B(0) \sim \frac{k\frac{\psi^2}{2}\delta_{AB}}{z^2} + \frac{if_{AB}{}^C J_C(0)}{z},
\end{equation}
where $\delta_{AB}$ is the Killing form on $\mathfrak{g}$, and $\psi^2$ is the squared length of the long root of $\mathfrak{g}$. Denoting $[J_A]\in\zhu_{s}(V)$ as $j_A$ and using \eqref{products}, it is straightforward to compute products in the Zhu algebra,
\begin{equation}
j_A \star j_B = (j_A j_B) + i\hbar f_{AB}{}^C j_C,
\end{equation}
where $(j_A j_B)=[(J_A J_B)]$ is the 2d conformal normal ordering. The stress-energy tensor $[T]=t$ is an element of the center of $\zhu_{s}(V)$, which is not independent due to the Sugawara construction,
\begin{equation}
T = \frac1{\psi^2(k+h^\vee)}\sum_{A,B} \delta^{AB}(J_A J_B).
\end{equation}
Now we determine the trace using \eqref{twtr_formula}. For the stress tensor $t$, we have:
\begin{equation}
T_s(t)  = \hbar^2 \frac{\trace_{\widetilde{M}_{\rm min}^{(0)}}\left\{s_{\widetilde{M}_{\rm min}^{(0)}}(-1)^F  L_0\right\}}{\trace_{\widetilde{M}_{\rm min}^{(0)}}\left\{s_{\widetilde{M}_{\rm min}^{(0)}}(-1)^F \right\}} = \hbar^2\widetilde{\Delta}_{\rm min},
\end{equation}
where we used $o_{s}(T)=L_0$, which simply measures the dimension $\widetilde{\Delta}_{\rm min}$ of the module of minimal dimension. Recall that $T(z)$ originates from the spinning Schur operator and must disappear from the cohomology in the 3d limit. Also notice that $t$ does not mix with the dimension one operators (\emph{i.e.}, with currents $j_A$) for symmetry reasons, while the above equation manifests mixing with the identity. Altogether this implies:
\begin{equation}
t - \hbar^2 \widetilde{\Delta}_{\rm min} \in N.
\end{equation}
In the rank one case, the null ideal $N$ is in fact generated by $t-\hbar^2\widetilde{\Delta}_{\rm min}$. When $\mathfrak{g}$ has higher rank, further generators (corresponding to Joseph relations) might appear.

Now let us look at $(j_A j_B) = [(J_A J_B)]$. This operator can mix both with the identity and with $j_A$. It follows from the symmetry that $T_s( (j_A j_B)) \propto \delta_{AB}$, so clearly by the Sugawara relation,
\begin{equation}
T_s( (j_A j_B) ) = \frac{\psi^2(k+h^\vee)}{\dim\mathfrak{g}} T_s(t) \delta_{AB} = \frac{\psi^2(k+h^\vee)}{\dim\mathfrak{g}} \hbar^2\widetilde{\Delta}_{\rm min} \delta_{AB},
\end{equation}
which determines mixing with the identity. To clarify mixing with $j_A$, we compute the following:
\begin{align}
\label{mixing21}
T_s \left( (j_A j_B)\star j_C \right)= T_s(j_A\star j_B\star j_C) - i\hbar f_{AB}{}^D T_s(j_D\star j_C).
\end{align}
Rewrite the first term as
\begin{align}
T_s(j_A\star j_B\star j_C)=\hbar^3 \frac{\trace_{\widetilde{M}_{\rm min}^{(0)}}\left\{s_{\widetilde{M}_{\rm min}^{(0)}}(-1)^F  J_A^0 J_B^0 J_C^0\right\}}{\trace_{\widetilde{M}_{\rm min}^{(0)}}\left\{s_{\widetilde{M}_{\rm min}^{(0)}}(-1)^F \right\}} = x f_{ABC},
\end{align}
where for convenience we wrote $o_s(J_A)=J_A^0$ for the zero mode, and we also anticipated that the answer must be proportional to $f_{ABC}$ for symmetry reasons. Let us antisymmetrize in $A,B$, so that under the trace we get a commutator $[J_A^0, J_B^0]=if_{AB}{}^C J_C^0$, and on the right we simply get $xf_{ABC}-xf_{BAC}=2xf_{ABC}$, therefore obtaining:
\begin{align}
2x f_{ABC} &= if_{AB}{}^D \hbar^3 \frac{\trace_{\widetilde{M}_{\rm min}^{(0)}}\left\{s_{\widetilde{M}_{\rm min}^{(0)}}(-1)^F  J_D^0 J_C^0\right\}}{\trace_{\widetilde{M}_{\rm min}^{(0)}}\left\{s_{\widetilde{M}_{\rm min}^{(0)}}(-1)^F \right\}}= i\hbar f_{AB}{}^D T_s(j_D\star j_C)\cr 
&\Rightarrow T_s(j_A\star j_B\star j_C)=\frac{i\hbar}{2} f_{AB}{}^D T_s(j_D\star j_C).
\end{align} 
Inserting this into \eqref{mixing21}, we find
\begin{equation}
T_s \left( (j_A j_B)\star j_C \right)= - \frac{i\hbar}{2} f_{AB}{}^D T_s(j_D\star j_C),
\end{equation}
which describes the mixing of $(j_Aj_B)$ with $j_C$. Let us now subtract all mixing with the lower-dimensional operators and define $:j_A j_B:$, the 3d normal ordering, by the following equation,
\begin{equation}
:j_Aj_B: = (j_Aj_B) + \frac{i\hbar}{2}f_{AB}{}^C j_C - \frac{\psi^2(k+h^\vee)}{\dim\mathfrak{g}}\hbar^2 \widetilde{\Delta}_{\rm min} \delta_{AB}.
\end{equation}
It follows from the construction that the operator $:j_Aj_B:$ is orthogonal to all lower dimension operators, and the star product can be written as
\begin{equation}
j_A \star j_B = :j_A j_B: + \frac{i\hbar}{2} f_{AB}{}^C j_C +\frac{\hbar^2}{4} \mu\psi^2\delta_{AB},\quad \mu= \frac{4(k+h^\vee)\widetilde{\Delta}_{\rm min}}{\dim\mathfrak{g}}.
\end{equation}
We will check this prediction in \cite{DW} by a more direct computation. Notice that because we took quotient by the ideal containing $t-\hbar^2\widetilde{\Delta}_{\rm min}$, the relation $\sum_{A,B} \delta^{AB}:j_Aj_B:=0$ holds in this algebra.

\subsection{Relation to the $C_2$ algebra and Beem-Rastelli conjecture}
The so-called $C_2$-algebra $R_V$ of a VOA $V$ is another important algebraic object defined by Zhu \cite{Zhu}. It is constructed as a quotient of $V$ by the ideal of operators containing derivatives. More precisely, one considers a subspace $C_2(V)\subset V$ spanned by vectors of the form $a_{-2} b$, for all $a,b\in V$. Then $V/C_2(V)$ has the structure of the commutative Poisson superalgebra, with the product induced by $a_{-1}b$ and the Poisson bracket induced by $a_0 b$. 

There is a well-known relation of $R_V$ to $\zhu(V)$ (or $\zhu_{s}(V)$ in our case), see \cite[Proposition 2.17(c)]{DSK} \cite[Proposition 3.3]{ARAKAWA2014261}, \cite[Lemma 4.3]{2016arXiv160500138A}. Namely, if we consider the associated graded ${\rm gr }\zhu_{s}(V)$ (it corresponds to taking the limit $\hbar\to 0$ if we choose to keep $\hbar$ in formulas), which is a Poisson algebra itself, then there exists a surjective homomorphism of Poisson algebras
\begin{equation}
\label{RVmap}
\eta_V: R_V \twoheadrightarrow {\rm gr }\zhu_{s}(V).
\end{equation}
This fact is quite trivial, and we now sketch a proof. The property 6 mentioned above states that $L_{-1}a + \hbar L_0 a \in O_{s}(V)$, so $L_{-1}a \in O_{s}(V) + \hbar V$. If we view ${\rm gr }\zhu_{s}(V)$ as $\zhu_{s}(V)/(\hbar \zhu_{s}(V)) \cong V/(O_{s}(V) + \hbar V)$, we see that derivatives $L_{-1}a$ are quotiented out in ${\rm gr }\zhu_{s}(V)$. Because derivatives are precisely what is quotiented out in the definition of $R_V$ (and not more), this implies the above map and that it is only surjective. Because the multiplication and Poisson bracket are induced by $a_{-1}b$ and $a_0 b$ respectively in both algebras, this implies that it is a homomorphism of Poisson algebras.

Not much is known about the kernel of \eqref{RVmap} in general, but there are some results available in special cases: if $V$ admits a PBW basis, then \eqref{RVmap} is an isomorphism \cite[Theorem 4.8]{2016arXiv160500138A}; for affine VOAs at integrable levels associated to classical Lie algebras, \cite{FEIGIN2011130,Feigin:2009xu} prove that \eqref{RVmap} is an isomorphism as well; and finally the same is proved for parafermion VOAs associated with the $\widehat{\mathfrak{sl}}_2$ in \cite{ARAKAWA2014261}. In all these cases one may regard $\zhu(V)$ as a deformation (quantization) of $R_V$.

One should also recall that the Beem-Rastelli conjecture on the Higgs branch \cite{Beem:2017ooy} states that the quotient $R_V/\cN$, where $\cN$ is the ideal generated by all nilpotent elements, is the Higgs branch chiral ring of the 4d $\cN=2$ theory in the cases when VOA arises via the 4d/2d correspondence of \cite{Beem:2013sza}. Additionally, the considerations of this paper prove that the Higgs branch chiral ring is
\begin{equation}
{\rm gr} \left( \zhu_{s}(V)/N \right) \cong {\rm gr }\zhu_{s}(V) / {\rm gr }N,
\end{equation}
which is the ``classical limit'' of $\cA_H$. Thus, if the Beem-Rastelli conjecture is true, it must be that
\begin{equation}
R_V/\cN \cong {\rm gr }\zhu_{s}(V) / {\rm gr }N.
\end{equation}
On the other hand, because $\eta_V$ in \eqref{RVmap} is surjective, one may write
\begin{equation}
{\rm gr }\zhu_{s}(V) / {\rm gr }N \cong (R_V/\ker\eta_V)/ {\rm gr }N \cong R_V/ \eta_V^{-1}\left({\rm gr }N\right).
\end{equation}
In other words, it must be true that
\begin{equation}
\cN = \eta_V^{-1}\left({\rm gr }N\right).
\end{equation}
One consequence of this is that all elements of ${\rm gr }N$ must be nilpotent in ${\rm gr }\zhu_{s}(V)$, and furthermore ${\rm gr }N$ must be precisely the nilradical of ${\rm gr }\zhu_{s}(V)$. Then it follows that
\begin{equation}
{\rm Specm}\, {\rm gr }\zhu_{s}(V) \cong {\rm Specm}\, R_V,
\end{equation}
which is, incidentally, the Conjecture 1 from \cite{10.2307/24523343}. So on the one hand, the Beem-Rastelli conjecture implies Conjecture 1 from \cite{10.2307/24523343} for the corresponding class of VOAs arising from the 4d physics. On the other hand, if we could prove that the ideal ${\rm gr} N\subset {\rm gr}\zhu_{s}(V)$ corresponding to the null ideal $N\subset \zhu_{s}(V)$ for trace $T_{s}$ is precisely the nilradical of ${\rm gr}\zhu_{s}(V)$, then the Conjecture 1 from \cite{10.2307/24523343} would readily imply the Beem-Rastelli conjecture.

We hope that methods of this paper will help to shed some more light on this web of conjectures.

\section{Future directions}\label{conclude}

We have established an interesting connection between protected sectors in 4d and 3d SCFTs. This opens new directions for both physics and math investigations, which we now briefly review.

On the physics side, the construction described here obviously calls for examples and applications. In a companion paper \cite{DW}, we partly fulfill this task by testing our construction against a number of examples, and using it as a tool to answer other questions.

Also, it is important to understand what happens in the $a_{\rm 4d}\geq c_{\rm 4d}$ case. For such theories, the relation between the Schur index and the $S^3$ partition function is problematic, the construction of this paper does not apply directly, and it is not obvious at the moment how to modify it. Perhaps it is possible to somehow subtract the 4d degrees of freedom that make the $\beta\to 0$ limit ill-defined. Another possible direction would be to study the hemisphere $HS^3$ partition function and the $HS^3\times S^1$ geometry in 4d. In the latter case, the limit $\beta\to 0$ might be well-defined, similar to how the $HS^3$ partition function of ``bad'' 3d $\cN=4$ theories  is defined even when the full $S^3$ partition function is not \cite{Dedushenko:2017avn,Dedushenko:2018icp}.

Mathematically, it would be interesting to determine under what conditions on the VOA, the short star-products that we obtain are even and Hermitian, as well as clarify strong non-degeneracy of $T_s$. Furthermore, a crucial step in the derivation was the use of modularity in Section \ref{sec:modularity}. For general VOAs, this is conjectural and should be properly addressed. Also, one could hope that maybe there exists a characterization of the null ideal $N$ that is more intrinsic to the VOA.

On a different note, it was remarked in \cite{ERS1} that short star-products provide new interesting structures in representation theory. In Section 4.2 of \cite{ERS1}, in particular, they discuss how traces in highest-weight modules give short star-products. What we have here is essentially a VOA version of this, which leads to the following questions:
\begin{itemize}
	\item On the one hand, we could study high-temperature (\emph{i.e.} small complex structure) limits of torus correlators for more general VOA modules (NS sector, R sector, or even general twisted modules). We could study both regular correlators and ``twisted'' correlators, when the module $M$ is $h$-invariant, $h\in{\rm Aut}(V)$, and we insert $h_M$ under the trace. In all these cases, the high temperature limit of torus correlators would determine some twisted traces on the corresponding Zhu algebras, and by the theorem of \cite{ERS1}, short star products.\footnote{Physically, general non-vacuum modules would correspond to dimensional reduction of the 4d theory in the presence of surface defects. Taking ``twisted'' traces would correspond to twisted compactifications.}
	
	\item The construction of the previous bullet-point still involves the high-temperature limit, and so requires understanding the modular behavior of the VOAs involved in it. Despite an obvious difficulty, it also has an intriguing positive aspect as it suggests the relation of short star products to the modularity properties of VOAs.
	
	\item On the other hand, we cold avoid talking about the high-temperature limit, and define twisted traces on Zhu algebras by simply applying the formula \eqref{twtr_formula}. Indeed, if we have two commuting finite order automorphisms $g,h\in {\rm Aut}(V)$, and an $h$-invariant $g$-twisted module $M$, we can use it to define an $h$-twisted trace on $\zhu_g(V)$ by the formula
	\begin{equation}
	T([a]) = \frac{\trace_{M_0} h_M (-1)^F o_g^M(a)}{\trace_{M_0} h_M (-1)^F},
	\end{equation}
	where as before, $h_M\in {\rm End}(M)$ is such that $h_M Y^M(a,z) h_M^{-1}=Y^M(h(a),z)$. This trace on $\zhu_g(V)$ is constructed from the trace in the $\zhu_g(V)$-module $M_0$ (cf. Section 4.2 of \cite{ERS1}).
	
	\item Finally, both the above equation and the equation \eqref{twtr_formula} earlier in this paper look problematic when $M_0$ is infinite-dimensional. Nevertheless, the trace map $T$ might still be well-defined, infinities in the numerator and the denominator canceling each other (and more generally, if there are more than one $M_0$, infinities between different modules canceling, as explained in Remark \ref{modularity_remark}). We essentially ignored this subtlety in this paper, assuming that it could be resolved via a proper regularization. Indeed, as we check in examples in \cite{DW}, the answers that follow from the high-temperature limit approach always end up being correct (for the $c_{\rm 4d}> a_{\rm 4d}$ theories), even when it is known that the corresponding modules of weight $\widetilde{\Delta}_{\rm min}$ have infinite-dimensional weight subspaces. This point deserves to be addressed as well.
	
\end{itemize}

It is also intriguing to ask whether various systems of differential equations obeyed by torus correlators, such as the Knizhnik-Zamolodchikov-Bernard equations in the affine case \cite{Knizhnik:1984nr,Bernard:1987df,Etingof:1993bj,Felder:1994pb,1999math......8115E}, play any role. In particular, whether one can use them to make any general statements about the $\tau\to0$ limits of correlation functions.

\section*{Acknowledgements}

I thank M.~Fluder and Y.~Wang	for collaborations on related projects \cite{Dedushenko:2019yiw} and \cite{DW}. I also acknowledge useful conversations, comments, and/or correspondence with: T.~Arakawa, T.~Creutzig, P.~Etingof (whom I also thank for sending the draft of \cite{ERS1}), B.~Feigin, L.~Rastelli, D.~Simons-Duffin. I also thank the anonymous referee for useful suggestions. Significant part of this work was completed when I was a member at (and was supported by) the Walter Burke Institute for Theoretical Physics, with the additional support from the U.S. Department of Energy, Office of Science, Office of High Energy Physics, under Award No de-sc0011632, as well as the Sherman Fairchild Foundation.

\bibliographystyle{utphys}
\bibliography{VOA_and_DQ}

\end{document}